\documentclass[fleqn,usenatbib]{mnras}

\usepackage{newtxtext,newtxmath}

\usepackage[T1]{fontenc}
\usepackage{ae,aecompl}

\usepackage{graphicx}	

\newcommand {\bc}{\begin {center}}
\newcommand {\ec}{\end {center}}
\newcommand {\beq}{\begin {equation}}
\newcommand {\eeq}{\end {equation}}

\newcommand{\sgr}{Sgr\,A$^{*}$\,}

\newcommand {\Chandra}{{\it Chandra }}

\usepackage[usenames,dvipsnames]{xcolor}
\usepackage{tikz}

\def\c {{\rm c}}

\def\deg{$^{\circ}$}

\title[X-raying molecular clouds]
{X-raying molecular clouds with a short flare: \\ probing statistics of gas density and velocity fields}
\author[Khabibullin et al.]{
I.~Khabibullin$^{1,2}$, E.~Churazov$^{1,2}$, R.~Sunyaev$^{1,2}$, C.~Federrath$^{3}$, D.~Seifried$^{4}$, S.~Walch$^{4}$
\\ 
$^{1}$ MPI f\"ur Astrophysik, Karl-Schwarzschild str. 1, Garching D-85741, Germany\\
$^{2}$ Space Research Institute, Profsoyuznaya str. 84/32, Moscow, 117997, Russia\\
$^3$ Research School of Astronomy and Astrophysics, Australian National University, Canberra, ACT 2611, Australia\\
$^4$ I. Physikalisches Institut, Universit\"at zu K\"oln, Z\"ulpicher Str. 77, D-50937 K\"oln, Germany\\
}

\date{Accepted XXX. Received YYY; in original form ZZZ}

\pubyear{2019}

\begin{document}
\label{firstpage}
\pagerange{\pageref{firstpage}--\pageref{lastpage}}
\maketitle

\begin{abstract}
{
We take advantage of a set of molecular cloud simulations to demonstrate a possibility to uncover statistical properties of the gas density and velocity fields using reflected emission of a short (with duration much less than the cloud's light-crossing time) X-ray flare. Such situation is relevant for the Central Molecular Zone of our Galaxy where several clouds get illuminated by a $\sim110$ yr-old flare from the supermassive black hole Sgr~A*. Due to shortness of the flare ($\Delta t\lesssim1.6$~yrs), only a thin slice ($\Delta z\lesssim0.5$ pc) of the molecular gas contributes to the X-ray reflection signal at any given moment, and its surface brightness effectively probes the \textit{local} gas density. This allows reconstructing the density probability distribution function over a broad range of scales with virtually no influence of attenuation, chemo-dynamical biases and projection effects. Such measurement is key to understanding the structure and star-formation potential of the clouds evolving under extreme conditions in the CMZ. {For cloud} parameters similar to the currently brightest {in X-ray reflection molecular complex Sgr~A}, the sensitivity level of the best available data is sufficient only for marginal distinction between solenoidal and compressive forcing of turbulence. Future-generation X-ray observatories with large effective area and high spectral resolution will dramatically improve on that by minimising systematic uncertainties due to contaminating signals. Furthermore, measurement of the iron fluorescent line centroid with sub-eV accuracy in combination with the data on molecular line emission will allow direct investigation of the gas velocity field.
}
\end{abstract}
%



\section{Introduction}
\label{s:introduction}

{
~~~~~~~Reflection of X-ray emission on molecular clouds in the Central Molecular Zone (CMZ) has proved to be a powerful tool for reconstructing {the activity} record of the Milky Way's supermassive black hole (SMBH) \sgr over past few hundred years \citep[][for a review]{1993ApJ...407..606S,1996PASJ...48..249K,2004A&A...425L..49R,2007ApJ...656L..69M,2010ApJ...714..732P,2010ApJ...719..143T,2012A&A...545A..35C,2013PASJ...65...33R,2015ApJ...815..132Z,2017MNRAS.468.2822K,2018A&A...610A..34C,2019MNRAS.484.1627K,2013ASSP...34..331P}. Namely, it has been shown that \sgr experienced at least one flare $\sim 110$ yrs ago, which had { a total fluence} of $\sim10^{47}$ erg and lasted for not longer than $\Delta t\lesssim 1.6$ yrs \citep{2017MNRAS.465...45C,2018A&A...612A.102T}. The latter point follows directly from rapid variability of the X-ray emission observed from the smallest substructures inside the illuminated molecular clouds \citep{2013A&A...558A..32C,2017MNRAS.465...45C}, and it has several important implications.
}

{
First of all, this provides strong support for the X-ray reflection paradigm itself, meaning that {the bulk} of the observed X-ray emission cannot be produced {due to interaction of molecular gas with cosmic rays \citep[either low energy protons or high energy electrons, e.g.][]{2002ApJ...568L.121Y,2009PASJ...61..901D,2012A&A...546A..88T,2018ApJ...863...85C}}.
}

{
Second, it allows to limit possible scenarios for the primary source of illumination \citep[e.g.][]{2012MNRAS.421.1315Z,2019MNRAS.486.1833S}, since the minimum required luminosity exceeds $10^{39}$ erg/s for $\Delta t=1.6$ yrs, while it equals few$\times 10^{44}$ erg/s, i.e. Eddington luminosity of \sgr, for $\Delta t$ as short as 1 hr \citep{2017MNRAS.465...45C}.  Further investigation of the light curves of the smallest resolved substructures will help improving on that even more \citep{2017MNRAS.471.3293C}.
}

\begin{figure}  
\begin{center}
\includegraphics[viewport=30 210 590 650,width=1\columnwidth]{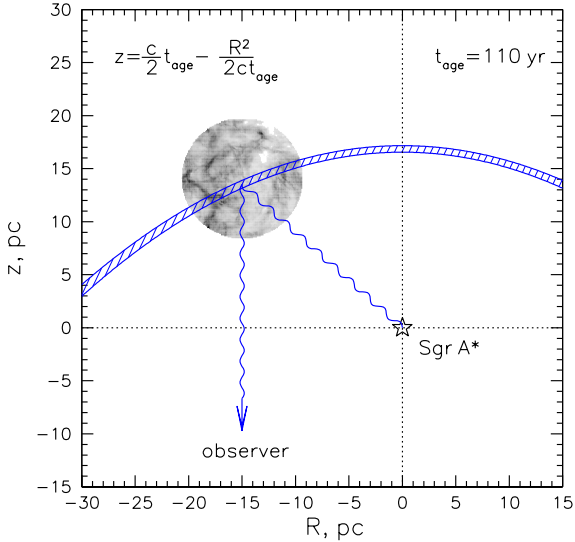}
\caption{
{
Geometry of reflection for a short Sgr\,A* flare (happened $t_{\rm age}=110$ years ago) on a molecular cloud, viewed from above the Galactic Plane, $z$ is the line-of-sight distance and $R$ is the projected distance from Sgr\,A*. The light propagates from the source to the cloud and then {to the observer}, so that equal-time-delay requirement ensures that only the gas between two parabolas $z=\frac{c}{2}t_{\rm age}-\frac{R^2}{2ct_{\rm age}}$ (hatched region) contributes to the reflected emission viewed by a distant observed at $z=-\infty$.
}
}
\label{f:parabola}
\end{center}
\end{figure}

\begin{figure}  
\begin{center}
\includegraphics[viewport= 30 90 440 370, width=1\columnwidth]{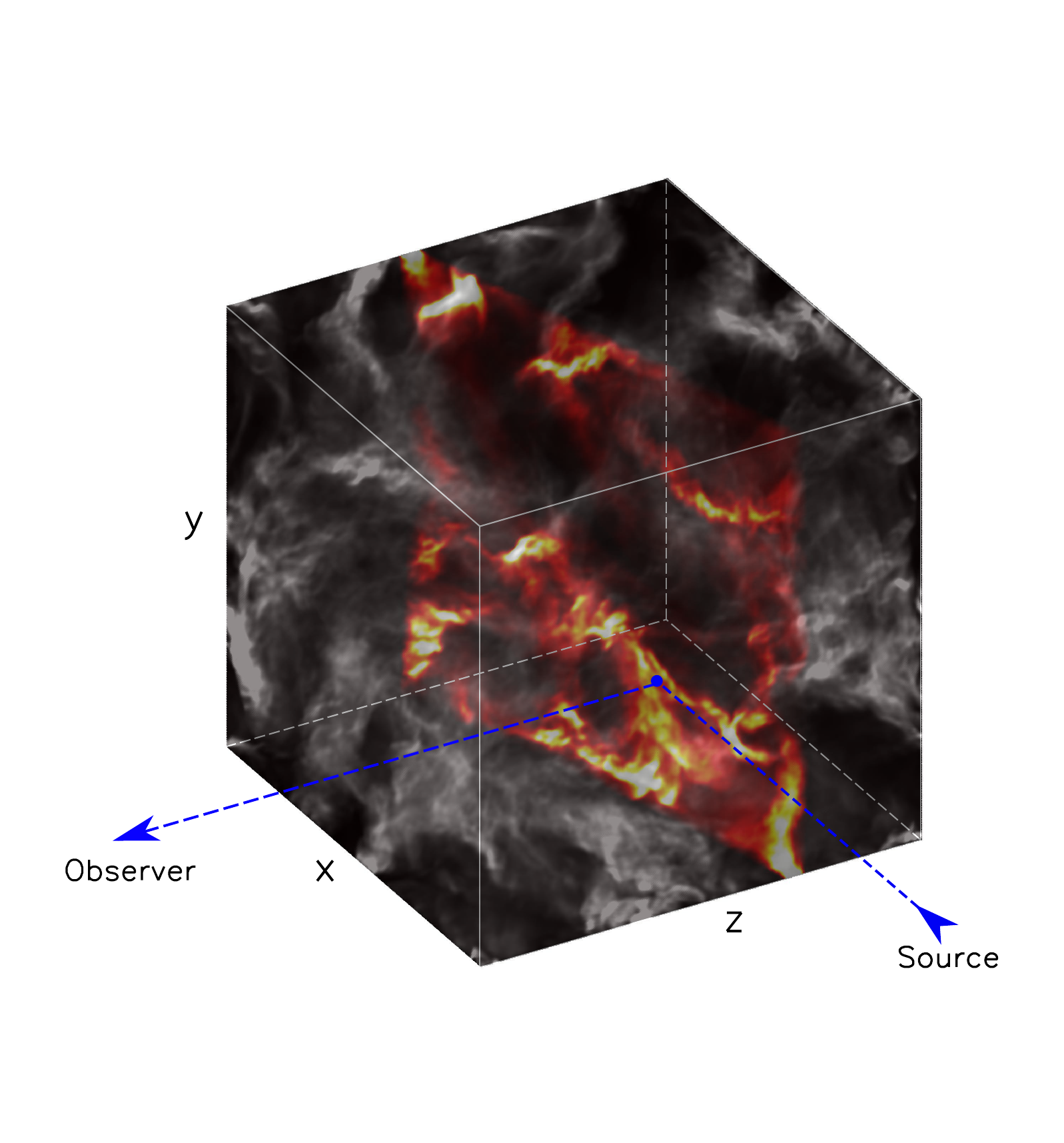}
\caption{
{
Schematic illustration of molecular cloud illumination by a short X-ray flare. At any given moment, a thin slice of the gas (shown in bright colour) is illuminated and contributes to the reflected emission visible to a distant observer. {A weaker signal associated with the second scattering is discussed in \S\ref{ss:second}.} All the intervening gas along the {photons' path} from the source to the reflector and to the observer (blue dashed lines) contributes to { the attenuation} of the reflection signal, {which is set by the column density of the intervening gas and effective (scattering and absorbing) cross section in a given energy band}.
}
}
\label{f:sketch}
\end{center}
\end{figure}

{
Finally, short duration of the flare implies that only a very thin layer of molecular gas, {$\delta z\sim \varv_{z}\Delta t\lesssim 0.5\,({\varv_z}/{c})\,(\Delta t/ 1.6\,\mathrm{yrs})$ pc}, contributes to the reflection signal at any given moment {(see Fig. \ref{f:parabola} and \ref{f:sketch} for illustrations)}, where the line-of-sight speed of the illumination front\footnote{{Hereafter by `illumination front' we mean the region contributing to the reflected emission as viewed by a distant observer, not the physical illumination front, which, of course, has spherical shape { centered} on the primary source and propagates { at} the speed of light.}} propagation $\varv_z$ ranges from $0.5c$ to $\infty$ depending on the relative position of the primary source, cloud and observer \citep[e.g.][]{2017MNRAS.465...45C}. As far as $ \varv_z$ does not exceed $c$ by a large factor, the thickness of the layer stays much smaller than the size of the whole molecular cloud, $L\sim10$ pc, so it is the local number density rather than projected column density of molecular gas that determines the observed X-ray emission.}

As a result, density fluctuations inherent to the molecular gas imprint themselves both in the spatial fluctuations of the reflected emission and temporal variation of the signal at any given position due to propagation of the illumination front across the cloud. Over the time-span of available \textit{Chandra} and \textit{XMM-Newton} data, $\Delta t\sim 20$ years, a $\sim \varv_z\Delta t\sim 6 ({\varv_z}/{c})$ pc-thick slab of the molecular gas has been 'scanned' by the X-ray flare. Comparison of statistical properties of temporal variations over this period with those of spatial variations derived from individual 'snapshots' (under assumption of fluctuations isotropy) allowed \citet{2017MNRAS.465...45C} to measure the illumination front propagation speed, $ v_{z}\approx0.7c$, and derive relative line-of-sight position, $z\approx10$ pc behind \sgr, {for the molecular complex Sgr~A} which is currently the brightest one in X-ray reflected emission \citep{2010ApJ...714..732P,2013A&A...558A..32C,2018A&A...612A.102T}. 

Moreover, high penetrative power of X-rays above 3 keV and very weak sensitivity of the reflection albedo to {the thermodynamic} and chemical state of gas imply that the observed signal should be a linear and pretty much unbiased proxy for the local gas density inside the illuminated layer. Thinness of this layer means {virtually no averaging} due to line-of-sight projection is involved, so the statistics of X-ray surface brightness $I_{X}$ {is directly linked to} statistics of the total molecular gas density $\rho$, namely $\rho\propto I_{X}$ {as long as the } absorbing column density is below few$\times 10^{23}$ cm$^{-2}$ \citep{2017MNRAS.468..165C}.  This is in drastic contrast to the commonly used line emission proxies which are hindered by projection effects, self-shielding, excitation- and chemistry-related biases \citep[e.g.][and references therein]{2017ApJ...835...76M}.         

The simplest statistics of the gas density is its Probability Distribution Function $PDF(\rho)\propto\frac{dV(\rho)}{d\rho}$, i.e. the fraction of volume filled with the gas of given density $\rho$. Since {the density field} inside molecular clouds is shaped {by a complex} interplay between supersonic turbulence, self-gravity, magnetic fields and stellar feedback, the gas density $PDF$ bears invaluable information on how this interplay actually proceeds and in which way it determines the dynamical state of the cloud \citep[][]{2014ApJ...781...91G,2014Sci...344..183K,2017MNRAS.466..914D,2018ApJ...859..162C,2018MNRAS.474.5588D}.

Namely, quasi-isothermal supersonic MHD turbulence is believed to define the density field at $1-10$ pc scales, resulting in the log-normal distribution function \citep{1994ApJ...423..681V} with the width determined by the Mach number of the turbulent motions, nature of forcing and {relative magnetization of the gas} \citep{1997MNRAS.288..145P,2008ApJ...688L..79F,2012MNRAS.423.2680M,2015MNRAS.451.1380N,2015MNRAS.448.3297F}.
    
{ Self-gravity of the dense gas is expected to lead to the formation of a high-density tail at subparsec scales {\citep{2000ApJ...535..869K,2009A&A...508L..35K,2011ApJ...727L..20K,2013ApJ...763...51F,2014ApJ...781...91G}}, where the transition from turbulent to coherent motions {happens}, and the {large-scale} turbulent cascade links with the gravitationally-bound star-forming cores \citep[e.g.][]{Goodman1998}. Due to this, $\sim 0.1$ pc scales are crucial for determining a cloud's overall star-formation efficiency \citep[e.g.][]{Williams2000,Ward2007}.} {This 0.1 pc scale also coincides with the sonic scale \citep[e.g.][]{2016MNRAS.457..375F} and the typical width of interstellar filaments \citep{2011A&A...529L...6A}, important building blocks of molecular clouds and the primary sites for star formation.}

{ In this regard, dense gas of the CMZ presents a particularly interesting study case because of the very dynamic and extreme environment of the Galactic Center, and also because of low star formation efficiency inferred for it \citep[compared to the average efficiency measured for molecular clouds in the Galactic disk,][]{2013MNRAS.429..987L,2014MNRAS.440.3370K,2016A&A...586A..50G,2016arXiv161003499K, 2017MNRAS.469.2263B}.} {High density and pressure of the surrounding medium, high energy density of cosmic rays and intensive disturbances due to supernova explosions are all capable of affecting physics of molecular clouds in the CMZ, plausibly making them more similar to the molecular gas in high-redshift star-forming galaxies \citep[e.g.][]{2013MNRAS.435.2598K}. }

Measuring the $PDF(\rho)$ {over a broad} range of the scales is a key for clarifying the particular mechanism(s) responsible for this suppression, e.g. high level of solenoidally-forced turbulence \citep{2016ApJ...832..143F}, and also whether this is a generic property of such an environment {or whether it differs from cloud} to cloud, e.g. as a result of the different orbital evolution histories \citep{2016arXiv161003502K}. {The evolution of the dense gas flows in centres of galaxies is very interesting from { a theoretical} point of view \citep[e.g.,][]{2018MNRAS.475.2383S}, while it is also of great practical importance in connection with feeding (and feedback) of supermassive black holes \citep[e.g.,][]{2012NewAR..56...93A}.}

It turns out that both duration of the Sgr A* X-ray flare and angular resolution {accessible with current X-ray observatories} ($\sim 1$ arcsec, corresponding to $\sim$0.04~pc at the Galactic Center distance) in principle do allow reconstruction of the gas density $PDF$ from $\sim10$ pc down to the subparsec scales based on the maps of reflected X-ray emission, and the first attempt of such study has been performed by {\citet{2017MNRAS.471.3293C}}.  The observed $PDF(I_X)$, based on \Chandra data taken in 2015, appears to be well described by the log-normal distribution with width $\sigma_I\approx 0.8$, where $\sigma_I$ is the standard deviation of the surface brightness normalised by its mean. Taken {at face} value, this would imply {\citep[using the velocity dispersion inferred from molecular line emission, e.g.][]{2010ApJ...714..732P}} predominantly solenoidal driving of turbulence inside this cloud, similar to the conclusion drawn by \cite{2016ApJ...832..143F} after analysing properties of another CMZ cloud, the so-called ``Brick'' cloud, which is indeed characterised by very low star formation efficiency  \citep{2016arXiv161003499K,2016arXiv161003502K,2017MNRAS.469.2263B}.

However, the observed shape of $PDF(I_X)$ is subject to several statistical distortions, both at the low and high surface brightness (viz. low and high density) ends of the distribution. This severely limits the dynamic range of the scales over which a reliable measurement can be done, so that the least affected part of the $PDF$ spans only a range of factor of $\sim 5$ in probed densities. The impact of some of these distortions has been illustrated by \citet{2017MNRAS.471.3293C} on analytically-constructed examples of the molecular gas density distributions.  

In the current study, we take advantage of a set of {idealised simulations of supersonic isothermal turbulence} \citep{2008ApJ...688L..79F} and global simulations of molecular clouds {\citep[from the SILCC-Zoom project{; see}][]{2015MNRAS.454..238W,2016MNRAS.456.3432G,2017MNRAS.472.4797S}} in order to explore the possibility to recover statistical properties of the gas density distribution and derive corresponding characteristics of the underlying turbulent velocity field. We discuss an optimal technique and formulate requirements for the observations with current and future {X-ray facilities} that will allow such {measurements} to be performed. Additionally, we discuss the feasibility and implications of the gas velocity field exploration via spectral mapping {using the} 6.4 keV iron fluorescent line.      

{
The structure of the paper reads as follows:  in Section \ref{s:simulations}, we outline the link between the gas density statistics and underlying fundamental processes operating inside molecular clouds (primarily supersonic turbulence) and describe sets of idealised and more realistic numerical simulations which we exploit for illustration of the density statistics reconstruction using illumination by short X-ray flares.
In Section \ref{s:reflection} we describe the basic properties of the X-ray reflection signal and its connection to the physical characteristics of the illuminated gas under conditions relevant for Sgr A$^*$ illumination of the clouds in the Central Molecular Zone.
In Section \ref{s:results}, we check whether { the dynamical} range provided by X-ray reflection signal allows robust diagnostics of the turbulence characteristics, given { the limitations} by opacity and finite duration of the flare. We consider observational limitations of the real data, formulate the requirements and discuss prospects of current and future X-ray facilities in Section \ref{s:discussion}. Summary of the paper follows in Section \ref{s:conclusions}.
}

\section{Gas density and velocity fields inside molecular clouds}
\label{s:simulations}
%

~~~~Molecular clouds are believed to arise due to cold gas condensation accompanied by gravitational contraction and fragmentation into dense collapsing cores, followed by seeding of pre-stellar objects, star formation and its feedback \citep{2007ARA&A..45..565M,2015ARA&A..53..583H}. Thus, they are extremely important building blocks of the interstellar medium, facilitating a complicated multi-scale link between global (galaxy-scale) gas flows and life-cycles (formation, evolution and death) of individual stars \citep{2015MNRAS.454..238W}. This fact is also reflected in the complexity of their internal structure, composed by hierarchies of (dynamic) substructures, shaped by various processes at different scales \citep{2012MNRAS.424.2599C,2017MNRAS.472.4797S}.  

\subsection{Supersonic turbulence}
\label{ss:turbulence}

At the largest scales ($\sim$1-10 pc), supersonic turbulence {is a key process} that determines properties of the gas density and velocity fields. For gas {with isothermal equation of state} (which is a good approximation for molecular gas as far as radiative cooling is efficient enough to mitigate heating due to kinetic energy dissipation), both analytic arguments \citep{1994ApJ...423..681V,1997MNRAS.288..145P} and numerical simulations \citep{1998ApJ...508L..99S,1999ApJ...513..259O,1999ApJ...524..169M,2001ApJ...546..980O,2007ApJ...665..416K} predict formation of a log-normal distribution of gas number density,
\begin{equation}
PDF(\rho)=p_{\rho}=\frac{dV(\rho)/V_0}{d\rho/\rho_0},
\end{equation}
which is essentially {a differential} distribution of the space volume (normalised by the cloud's total volume $V_0$) over the gas density $\rho$  (normalised by some mean density $\rho_0$) that is filling it.

Having defined $ s=\ln (\rho/\rho_0)$, in the case of {a log-normal} distribution one has {\citep[e.g.,][]{1997MNRAS.288..145P}}
\begin{equation}
\label{eq:lognorm_s}
PDF(s)=p_{s}=\frac{1}{\sqrt{2\pi\sigma_s^2}}\exp \left[-\frac{(s-s_0)^2}{2\sigma_s^2}\right],
\end{equation}
where $\sigma_s$ is the standard deviation of the logarithmic density, $ s_0=-\sigma_s^2/2$, and $p_s ds=p_{\rho}d\rho$ {by definition}. 

The width of the distribution $\sigma_s$ has been found to be set primarily by the Mach number of turbulent motions \citep{1997MNRAS.288..145P,1998PhRvE..58.4501P,2007ApJ...658..423K,2008ApJ...688L..79F,2012MNRAS.423.2680M} as 
\begin{equation}
\label{eq:sigmas}
\sigma_s^2=\ln(1+b^2\mathcal{M}_\beta^2)
\end{equation}
where $b$ is a proportionality coefficient and the Mach number $\mathcal{M}_\beta=\mathcal{M}/\sqrt{1+1/\beta}$ differs from the {hydrodynamic Mach number} $ \mathcal{M}=\varv/c_{s}$, $c_s=\sqrt{dP/d\rho}$, due to pressure contribution provided by turbulent magnetic fields at level $1/\beta=B^2/8\pi P$, with $P$ being the thermal gas pressure.

The scaling coefficient $b$ has been shown to depend on the nature of turbulence forcing, being equal to $1/3$ for solenoidal (divergence-free) forcing, $1$ for compressive (curl-free) one, and some intermediate value for a mixture of both \citep{2008ApJ...688L..79F}. {Although there} might be additional factors influencing the exact value of the parameter $b$  \citep[][e.g. it might depend on whether the gas is indeed isothermal or not]{2015MNRAS.451.1380N},  a factor of 3 difference between solenoidal and compressive forcing suggests that it might be in principle possible to distinguish between these two cases by solving for Eq.\ref{eq:sigmas} with $\sigma_s$ and $\mathcal{M}_\beta$ provided by observations \citep{2010A&A...512A..81F}.

\begin{figure}  
\begin{center}
\includegraphics[viewport= 30 200 590 650,width=1\columnwidth]{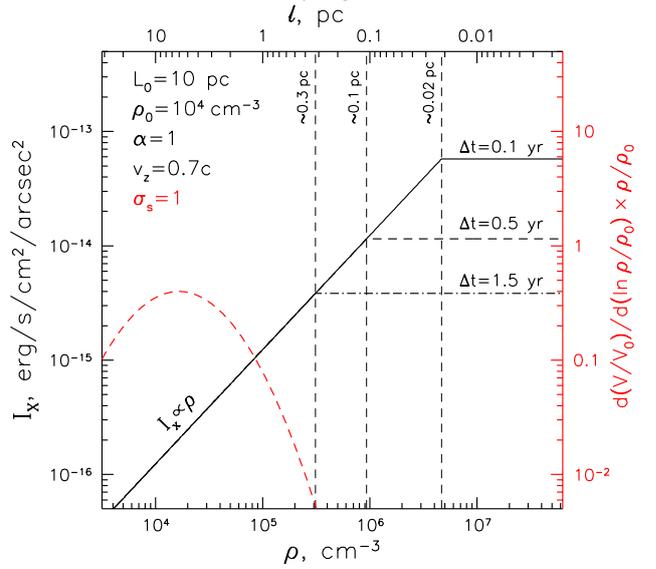}
\caption{Relation between surface brightness of reflected X-ray emission and gas density (and characteristic size) of a substructure inside a molecular cloud with power law {density (bottom} axis)-size(top axis) scaling relation $ \rho\propto l^{-\alpha} $ with $\alpha=1$, and $\rho_0=10^4$ cm$^{-3}$ at $L_0=10$ pc. Surface brightness normalisation corresponds to the flare fluence $ L_{X}\Delta t = 3\times 10^{46}$ erg (4-8 keV) and source-to-cloud distance $D_{sc}=30$ pc. {The front} propagation speed along the line-of-sight is set equal to $\varv_z=0.7c$, while the flare's duration is $ \Delta t=0.1,0.5 $ and 1.5 yr for solid, dashed and dash-dotted lines respectively. {The relation is linear down to the scale marked with thin vertical dashed lines for each flare's duration.} The red dashed curve (corresponding to the right y-axis) shows mass {(i.e. fraction of volume multiplied by density)} distribution between the scales for a cloud having log-normal density distribution with mean density $\rho_0=10^4$ cm$^{-3}$ and $\sigma_s=1$ (see text for details).
\label{f:tf1}
}
\end{center}
\end{figure}

In {Fig. \ref{f:tf1}}, we put a log-normal distribution of gas density in the context of the scales that are linearly probed by means of X-ray reflection. As can be seen, the dynamic range accessible to X-ray observations matches very well the range where the density $PDF(\rho)$ varies very prominently, meaning that the width of the density distribution $\sigma_s$ can be robustly inferred from the observed distribution of $I_{X}$. This picture is, however, based on a certain size-density relation, which is of course a gross simplification and we drop it by taking advantage of an extensive set of high-resolution (1024$^3$) 3D simulations of solenoidally or compressively forced  supersonic (with root-mean-square Mach number $\mathcal{M}\approx5.5$) {turbulence}\footnote{Available at \url{http://starformat.obspm.fr/starformat/project/TURB_BOX}.} \citep{2008ApJ...688L..79F}.

In depth analysis of these simulations at different evolutionary times {show} that the shape of $PDF(\rho) $ is indeed well described by the log-normal shape, and also that the mass-size scaling relation for the cloud's substructures is quite close to the prediction {of the third scaling relation of \citet{1981MNRAS.194..809L}} \citep[see, however,][for a discussion of possible caveats and level of uncertainty in that conclusion]{2011ApJ...727L..20K}. 

\subsection{Global picture}
\label{ss:global}

Of course, the supersonic turbulence is not the only agent shaping density and velocity fields inside molecular clouds. In particular, at smaller (sub-pc) scales, gravity starts playing an increasingly significant role, marking the transition from random turbulent motions to coherent ones at the so-called sonic scale $\lambda_{s}\sim0.1$ pc. Formation and contraction of compact gravitationally-bound substructures leads to a prominent distortion of the high end of the density distribution function, particularly in the form of power-law tails well above the extrapolation of the log-normal distribution to these scales \citep[e.g.,][]{1999ApJ...513..259O,2000ApJ...535..869K,2011MNRAS.410L...8C,2011ApJ...727L..20K,2013ApJ...763...51F,2014ApJ...781...91G,2017ApJ...834L...1B}. {It is worth mentioning, that the `cascading turbulence' paradigm bears certain caveats and recent observations \citep{2018MNRAS.477.2220T} seem to provide support to the scenario in which multi-scale collapse motions dominate in the turbulent velocity field \citep[e.g.][]{2018MNRAS.479.2112B}.}

This high density tail {is crucial} for the star formation rate and the star formation efficiency of the cloud {\citep[e.g.][]{2005ApJ...630..250K,2008ApJ...672.1006E,2011ApJ...743L..29H,2011ApJ...730...40P,2012ApJ...761..156F,2015ApJ...806L..36S,2018ApJ...863..118B}}. In this {respect}, molecular clouds in the CMZ are of particular interest because the star formation efficiency measured for them appears to be an order of magnitude lower compared to the molecular clouds in the Galactic disk \citep{2013MNRAS.429..987L,2014MNRAS.440.3370K,2016A&A...586A..50G,2016arXiv161003499K, 2017MNRAS.469.2263B}. This might be a consequence of the very specific {Galactic Center environment,} e.g. as a result of intensive forcing of solenoidal turbulence due to tidal shears {\citep{2016ApJ...832..143F}}, {or reflect some evolutionary track of the dense gas flows in the dynamic gravitational potential \citep{2015MNRAS.447.1059K,2018MNRAS.475.2383S,2019MNRAS.484.5734K,2019MNRAS.486.3307D}.}

Thus, it is clear that the simple picture considered earlier has relevance only {for a certain} range of scales, which is set by the evolutionary state and {environmental} conditions of a particular cloud.

In order to explore this in more detail, we take advantage of the output produced by 3D `zoom-in' simulations SILCC-Zoom \citep{2017MNRAS.472.4797S}, which are part of the SILCC (SImulating the LifeCycle of molecular Clouds) project {\citep{2015MNRAS.454..238W,2016MNRAS.456.3432G}}. 
{These simulations model the self-consistent formation and evolution of molecular clouds in their larger-scale galactic environment. They follow the non-equilibrium formation of H$_2$ and CO including (self-) shielding and important heating and cooling processes, which allows for an accurate description of the thermodynamical and chemical properties of the clouds. The clouds are modelled in a small section of a galactic disc with solar neighbourhood properties and a size of 500
pc $\times$ 500 pc $\times$ $\pm$ 5 kpc in which turbulence is initially driven by supernova explosions.}

{During the course of the simulations two 'zoom-in' regions are selected in which molecular clouds are about to form. The clouds in these zoom-in regions are then modelled with a high spatial resolution of {up to} 0.12 pc using adaptive mesh refinement while simultaneously keeping the galactic environment at lower resolution.}
Such a resolution of 0.12 pc {appears sufficient} for the flare duration $\gtrsim 0.3$ yrs, and it also translates into spatial binning at level {$\approx3$  arcsec} at the distance of the Galactic Center {\citep[$d_{GC}\approx 8.2$ kpc,][]{2019arXiv190405721A}}, which matches quite well the angular resolution provided by the best current (\textit{Chandra}) and future (\textit{Lynx} and \textit{ATHENA})  X-ray observatories (see Section \ref{ss:observe} for a discussion of the instrument resolution impact on probing the gas density $PDF$).

{A possible caveat of using these simulations is that molecular clouds {in the} CMZ might {have different} properties compared to the clouds in the Galactic disk, which were the primary focus of modelling in the SILCC project. {Nevertheless,} the properties {that} are of {greatest} importance for our study (i.e. sampling variance of the $PDF$ shape and impact of self-gravity and feedback on it) should be {qualitatively} similar, allowing to illustrate the possibility of the $PDF$ reconstruction over the turbulence-dominated range of scales. {Namely, the log-normal shape of the gas density $PDF$ (with a power-law tail at later times) is indeed recovered in these simulations when the dense gas is considered. Of course, these simulations also contain much more tenuous gas, so the full density $PDF$ has a more complicated shape reflecting multi-phase nature of the gas.} { Once again, we stress out that exact $PDF$ signatures of this multi-phase picture might be different for the CMZ clouds. }
}

In the next Section, we model the distribution function for the surface brightness of the reflected X-ray emission resulting from illumination of these simulated molecular clouds by a short X-ray flare. 

\section{Reflection of a short X-ray flare on molecular gas}
\label{s:reflection}

~~~~~ {The possibility} of using reflection of X-ray emission for simultaneous studies of interstellar medium and activity records of illuminating sources has been envisaged long ago \citep[e.g.][]{1980SvAL....6..353V}. Indeed, interaction of X-ray radiation with cold atomic and molecular gas is both very well understood from theoretical point of view \citep{1996AstL...22..648S,1998MNRAS.297.1279S,1998AstL...24..271V,1999AstL...25..199S} and in application to numerous astrophysical situations {\citep[see][for a recent review]{2010SSRv..157..167F}},{ including illumination of a secondary star in X-ray binaries {and reflection by its atmosphere} {\citep{1974A&A....31..249B}}, reflection on accretion disks and reprocessing of Active Galactic Nuclei emission on {the surrounding torus} {\citep[][]{1991MNRAS.249..352G}}.} Characteristics of the resulting emission (intensity, spectrum, polarization etc.) can be readily modelled, {especially when certain simplifying assumptions might be adopted.}

Namely, for the illumination of molecular clouds {in the} CMZ by the short X-ray flare from \sgr, one may safely neglect influence of X-ray radiation on the ionisation, thermal or chemical state of the irradiated gas \citep[see, however,][for possible consequences of powerful X-ray illumination over a more extended period of time]{2019MNRAS.486.1094M}. 

{Besides that, column densities of the molecular clouds {in the} CMZ typically do not exceed $N_{H}\,\sim 10^{24}$ cm$^{-2}$, hence their optical depth with respect to electron scattering, $\tau_{\rm T}\sim N_{H}\sigma_{\rm T} \lesssim 0.66$. Furthermore, the fraction of scattered to absorbed  photons scales as $\sigma_T/(\sigma_T+\sigma_{\rm ph})$, where $\sigma_{\rm ph}$ is cross-section for photoelectric absorption. Given that $ \sigma_{\rm ph}$ exceeds $\sigma_{\rm T}$ for photon energies $E\lesssim 15$ keV (for solar abundance of heavy elements), single-scattering approximation is sufficient in most cases, except for the most massive clouds like Sgr B2, the densest compact cores, and if Compton shoulder of the fluorescent line is of particular concern \citep[e.g.][]{2016A&A...589A..88M}. {The impact of the primary X-ray reflection signal contamination by doubly-scattered emission is described in Appendix \ref{s:appa}}.
}   

{ For a {known} geometry of illumination and observation, the problem then reduces to a single value describing {the efficiency} of X-ray reflection in a certain energy band, i.e. X-ray reflection albedo or effective scattering cross-section \citep[e.g.][]{2017MNRAS.471.3293C}. The latter should take into account the contribution of the fluorescent lines falling inside the energy band, and for {the 4-8 keV} band this results in {an effective} scattering cross-section $\sigma_{4-8}${, which is} a factor of 1.7 larger than $\sigma_{\rm T}$ for 90\deg -scattering and {for solar} abundance of heavy elements, as far as  $N_{H}\,\lesssim 3\times 10^{23}$ cm$^{-2}$\citep[see][for a detailed discussion]{2017MNRAS.471.3293C}. {These} conditions {ensure} that the illuminated gas is optically thin with respect to photoelectric absorption, so it fully contributes to the {reflected} signal. At $N_{H}\,\gtrsim 3\times 10^{23}$ cm$^{-2}$, some fraction of the gas gets self-shielded, and {the reflection} efficiency drops \citep{2017MNRAS.471.3293C}.}

{In the optically thin limit, the observed surface brightness of the reflected emission can be written as}
\begin{eqnarray}
\label{eq:ix}
I_X=\frac{L_X}{4\pi D_{sc}^2}\frac{\sigma_{4-8}}{4\pi}\,{\delta z}\,\bar{\rho}_H,
\end{eqnarray}
where $L_X$ is {the luminosity} of the illuminating source, $D_{sc}$ distance from it to the cloud in question, $\delta z =\varv_z\Delta t $ - thickness of the irradiated gas layer along the line-of-sight, and $\bar{\rho}_H$ is the average number density of the illuminated gas inside the aperture.

{For some fiducial values of the parameters entering it \citep[e.g.][]{2017MNRAS.471.3293C}, this equation predicts surface brightness in the 4-8 keV energy band equal to
\begin{eqnarray}
\label{eq:ixx}
I_X=1.2\times10^{-16}~\mathrm{erg~s^{-1}~cm^{-2}~arcsec^{-2}}~\times \nonumber\\
~\times~\left(\frac{L_X\Delta t}{3\times10^{46}~\mathrm{erg}}\right)\left(\frac{\varv_z}{0.7c}\right)
\left(\frac{D_{sc}}{30~\mathrm{pc}}\right)^{-2}\left(\frac{\bar{\rho}_H}{10^{4}\mathrm{cm^{-3}}}\right).
\end{eqnarray}}

As far as the size of the cloud $ L $ is {much} less than its distance to the primary source $ D_{sc}$, the illumination front propagation speed $\varv_z$ doesn't vary significantly across the whole extent of the illuminated region. In principle, it ranges from $0.5c$ to $\infty$, depending on the relative position of the primary source, cloud and observer, so that for the Sgr B2 cloud it was estimated at level of $3c$, while for the currently brightest in X-ray reflection {molecular complex Sgr A} it was measured equal to $0.7c$ \citep{2010ApJ...714..732P,2017MNRAS.465...45C}. This implies line-of-sight extent of the illuminated region at level {$\delta z= v_{z}\Delta t\approx 0.3 (\Delta t/\mathrm{yr})({\varv_z}/{c})$ pc}, which is significantly smaller than {the characteristic size of this molecular complex, $L\sim 10$ pc \citep{2010ApJ...714..732P},} {for $\Delta t \lesssim 2$ yrs and any $ \varv_z$ not exceeding $c$ by a large factor.}

The average number density of the illuminated gas inside some aperture of area $\Delta\Omega$ is given by 

\begin{equation}
\bar{\rho}_H(x,y,z)=\frac{1}{\delta z\Delta\Omega}\int_{\delta z} dz\int_{\Delta \Omega}{d\Omega}~ \rho_{H}(x,y,z),
\end{equation}
where $(x,y)$ are coordinates in {the plane of the sky} and $z$ is the line-of-sight coordinate. If {the size} of the aperture might be made {arbitrarily} small $ \Delta\Omega\rightarrow 0$, then $ \frac{1}{\Delta\Omega}\int_{\Delta \Omega}d\Omega\,\rho_{H}(x,y,z)\approx \rho_{H}(x,y,z)$, so that

\begin{equation}
\label{eq:rhoz}
\bar{\rho}_H(x,y,z)\approx \frac{1}{\delta z}\int_{\delta z} {dz}\,\rho_{H}(x,y,z).
\end{equation}

Of course, in reality {the smallest} possible $\Delta\Omega$ is set by {effective} spatial resolution of the X-ray data, determined both by angular resolution of the telescope and statistical significance of the signal, and we consider the impact of the corresponding limitations in Section \ref{ss:observe}.  

Let us consider a substructure inside the molecular cloud which has line-of-sight extent $l$ and constant density $\rho_{H,l}(x,y)$. If $l>\delta z$, than the integration in Eq.\ref{eq:rhoz} is trivial and gives $\bar{\rho}_H(x,y,z)=\rho_{H,l}(x,y)$. Combining this with Eq.\ref{eq:ix} one can see that $ I_{X}\propto \rho_{H,l}(x,y)$ in this case, so $ I_{X} $ simply maps the number density of such a substructure in the linear fashion.

Consider now the opposite case - a substructure of size  $l$ less than $\delta z$, i.e. having constant density $\rho_{H,l}(x,y)$ inside some subrange of $\delta z$ and zero density outside it. The integration in Eq.\ref{eq:rhoz} gives here $\bar{\rho}_H(x,y,z)=\frac{l}{\delta z}\rho_{H,l}(x,y)$. That means, one gets $ I_{X}\propto l \rho_{H,l}(x,y)$, {indicating} that reflected emission of smaller structures gets suppressed due to incomplete filling of the illumination front {thickness} by them. 

If there is a certain correspondence between the average density $\rho_{H,l}$ and size $l$ of a substructure, e.g. {in the form of a power law} scaling relation  $ \rho_{H,l}=\rho_{H}(l)\propto l^{-\alpha} $, one can combine the two cases considered above as:

\begin{eqnarray}
\label{eq:ix_prop}
I_X\propto \left\{\begin{array}{ll}
\rho_{H}(l), \,~~~~~~~~~~~~~ l>\delta z \\
\rho_{H}(l)^{1-1/\alpha}, \,~~~~~~ l<\delta z.
\end{array}
\right. 
\end{eqnarray}

For instance, {would the {third} scaling relation of \citet{1981MNRAS.194..809L}
 hold for the substructures},  $N_{H}(l)\equiv\rho_{H}(l)l={\rm const}$, hence $\rho_{H}(l)\propto 1/l$, i.e. $\alpha=1$ and $I_X\propto\rho_{H}(l)^0={\rm const}$ for $l<\delta z$. Clearly, a stronger ($\alpha>1$) or weaker ($\alpha<1$) scaling will result in, respectively, weaker or stronger suppression of X-ray reflection from substructures at {the high density} end.

{We illustrate this effect in Fig.~\ref{f:tf1} for various durations of the flare and a molecular cloud having size-density relation for substructures in the form of $\rho_{H}(l)\propto l^{-\alpha} $ with $ \alpha=1$, and gas {number density} $10^{4}$ cm$^{-3}$ at $l=10$ pc, {broadly similar to the currently brightest in X-ray reflection molecular complex Sgr A \citep[cf. Table 2 in][]{2010ApJ...714..732P} and the ``Brick'' molecular cloud \citep[][]{2016ApJ...832..143F}.} As can been seen, it is in principle possible to linearly probe scales at least down to 0.3 pc, corresponding to gas density $\sim 3\times10^{5}$ cm$^{-3}$ in this case. For flare duration as short as 0.5 yr (0.1 yr), the smallest linearly-probed scale is 0.1 pc (0.02 pc), corresponding to density $10^{6}$ cm$^{-3}$ ($5\times10^{6}$ cm$^{-3}$).} Thus, the dynamic range provided by X-ray reflection observations spans 1.5-2.5 orders of magnitude in density.

{It is worth {mentioning that} the integrated column density of such a cloud would be $ N_{H}\sim3\times 10^{23} $ cm$^{-2}$, while the scaling with $ \alpha=1$ means the column density stays the same over the whole range of substructure scales. This ensures applicability of the optically thin approximation over the full accessible dynamical range.} 

{In principle, it is possible that several substructures with size $ l\ll \delta z$ overlap in projection. The probability of such an occasion is determined essentially by their number density, intimately related to {the} gas density $PDF$. The expected shape (e.g., log-normal) of this distribution function decreases rapidly at the high density end (see Section \ref{s:simulations}), ensuring that the overlap probability is also a decreasing function of the gas density in this part of the distribution. Finite angular resolution of the real data {could}, of course, substantially boost this probability, but it also proportionally smears down {the X-ray reflection signal produced from such substructures}, so the resultant (i.e. summed and smeared) surface brightness is very unlikely {to overshoot significantly} the maximum $I_{X}$ level predicted by linear scaling {in Eq.}\ref{eq:ix_prop} (see Section \ref{ss:observe} for an in depth discussion of that point).}

Of course, real molecular clouds have much more complicated internal structure than described by a single density-size scaling relation for the substructures and gas density $PDF$ as we used for illustration above. For instance, smaller substructures might be preferentially embedded in bigger ones, and hence being subject to higher absorption and potentially to a certain degree of clustering. On the other hand, compact optically thick cores can `cast a shadow' {on all} the gas behind them, both from the point of view of the illuminating source and the observer. We shall check the impact of all these effects on the relation between the statistics of the X-ray reflection emission and the gas density field using outputs of numerical simulations of molecular clouds as described next.

\section{Results}
\label{s:results}
\subsection{Isothermal supersonic turbulence}
\label{s:stats}

\begin{figure*}
\includegraphics[width=0.63\textwidth,viewport= 20 260 570 600]{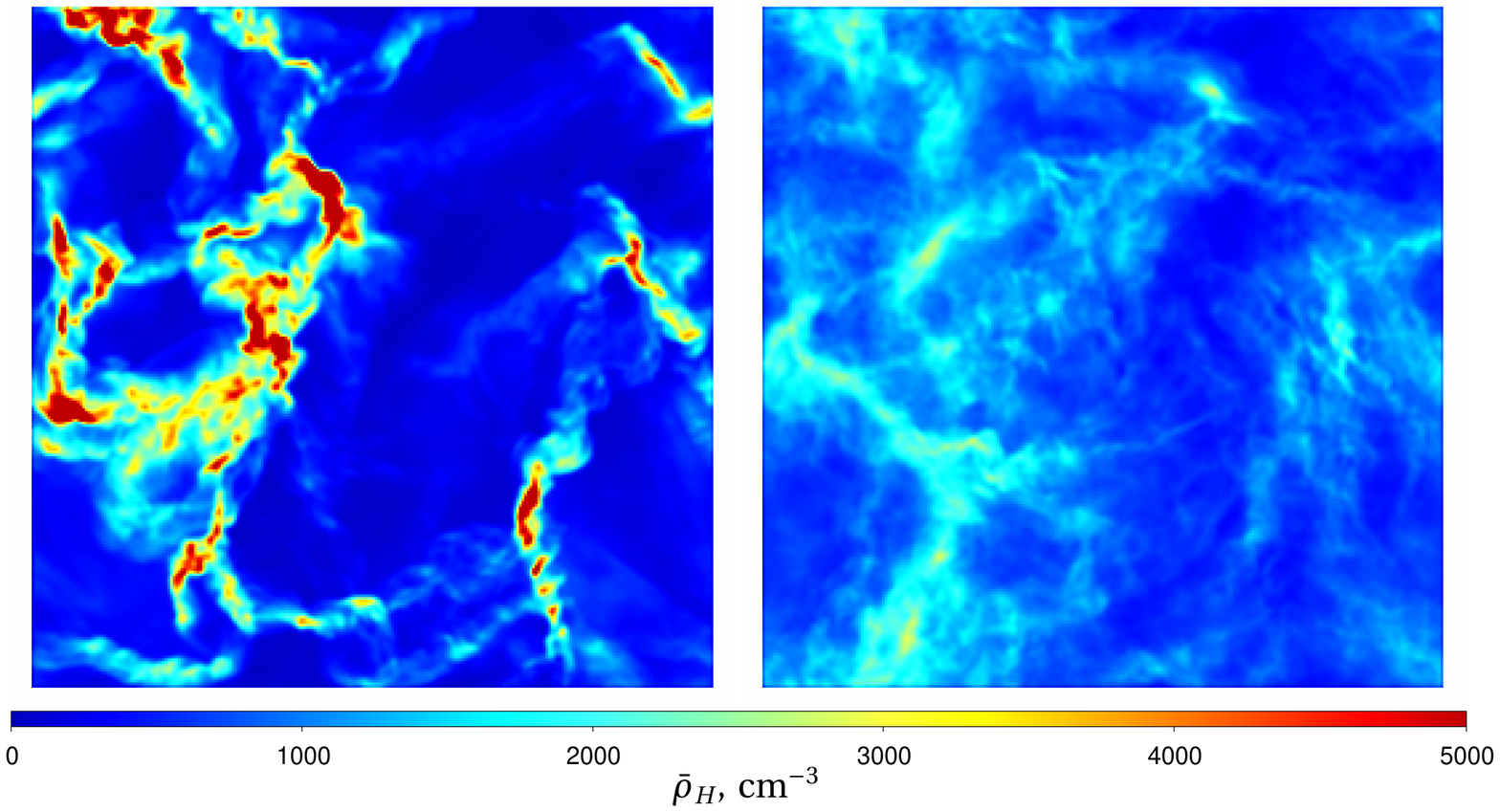}
\includegraphics[width=0.35\textwidth,viewport= 30 165 540 665]{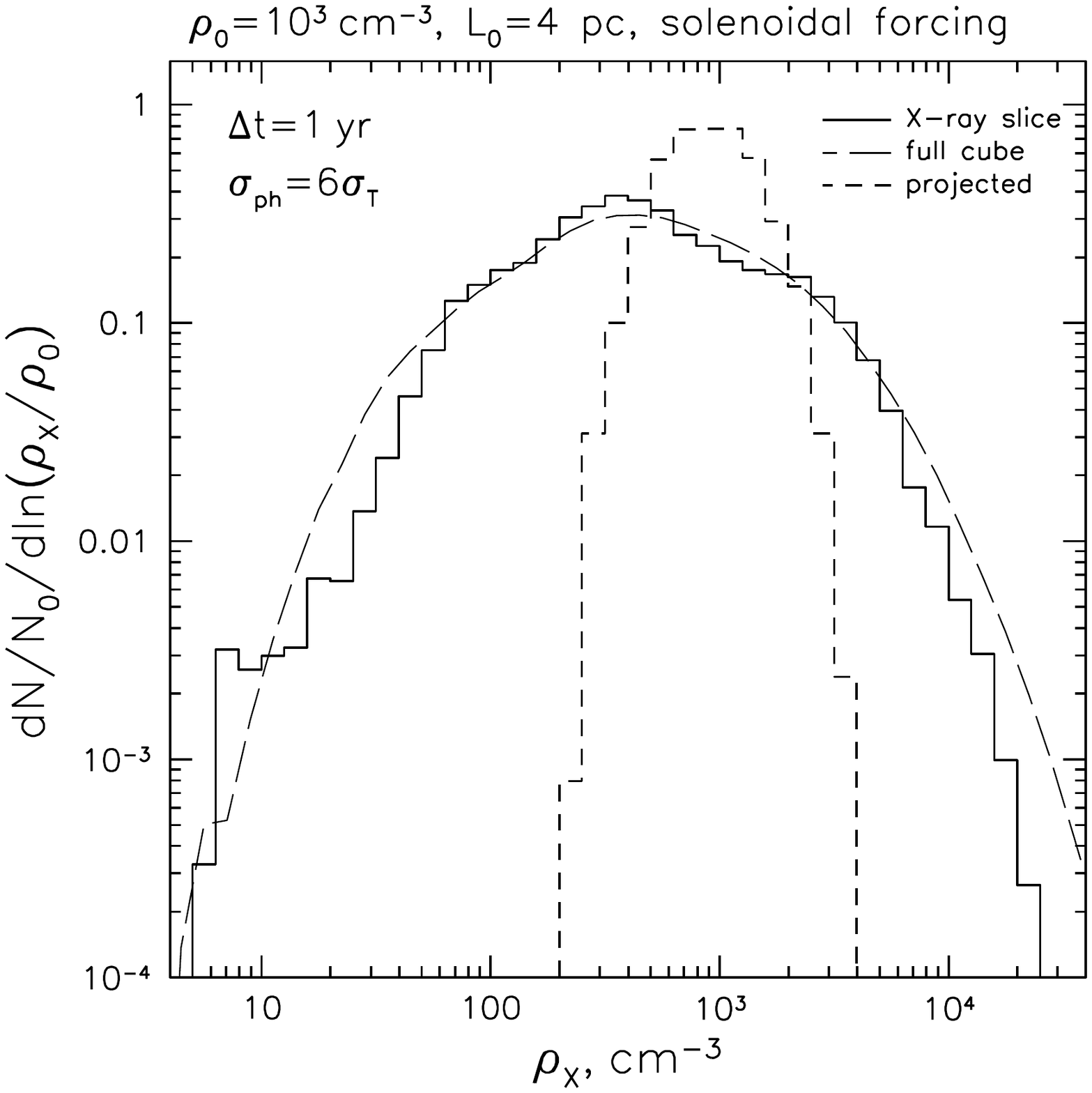}
\caption{Comparison of the gas density statistics as probed by X-ray reflection and column density techniques for a 4 pc box of simulated isothermal turbulence (solenoidal driving) with mean gas density $ \rho_0=10^{3}$ cm$^{-3}$. \textit{Left panel.} A map of the gas density {(cm$^{-3}$)} that would be revealed by an X-ray flare with duration $\Delta t=1$ yr passing through the middle of the cloud (see Fig.~\ref{f:sketch}). \textit{Middle panel.} A map of the gas density {(cm$^{-3}$)} averaged {by volume-weighted integration} over the whole line-of-sight axis of the simulation box.  \textit{Right panel.} Gas density \textit{PDF} as derived from the illuminated slice (solid line) compared to the \textit{PDF} of the line-of-sight-averaged (i.e. projected) density {(short-dashed line)}. The {long-dashed} curve shows gas density \textit{PDF} extracted from the whole box. Cross-section for X-ray attenuation was set equal to $ \sigma_{\rm ph}=6\sigma_{\rm T}$.}
\label{f:proj_fed_sol_c3_s6_t1}
\end{figure*}

\begin{figure*}
\includegraphics[width=0.63\textwidth,viewport= 20 260 570 600]{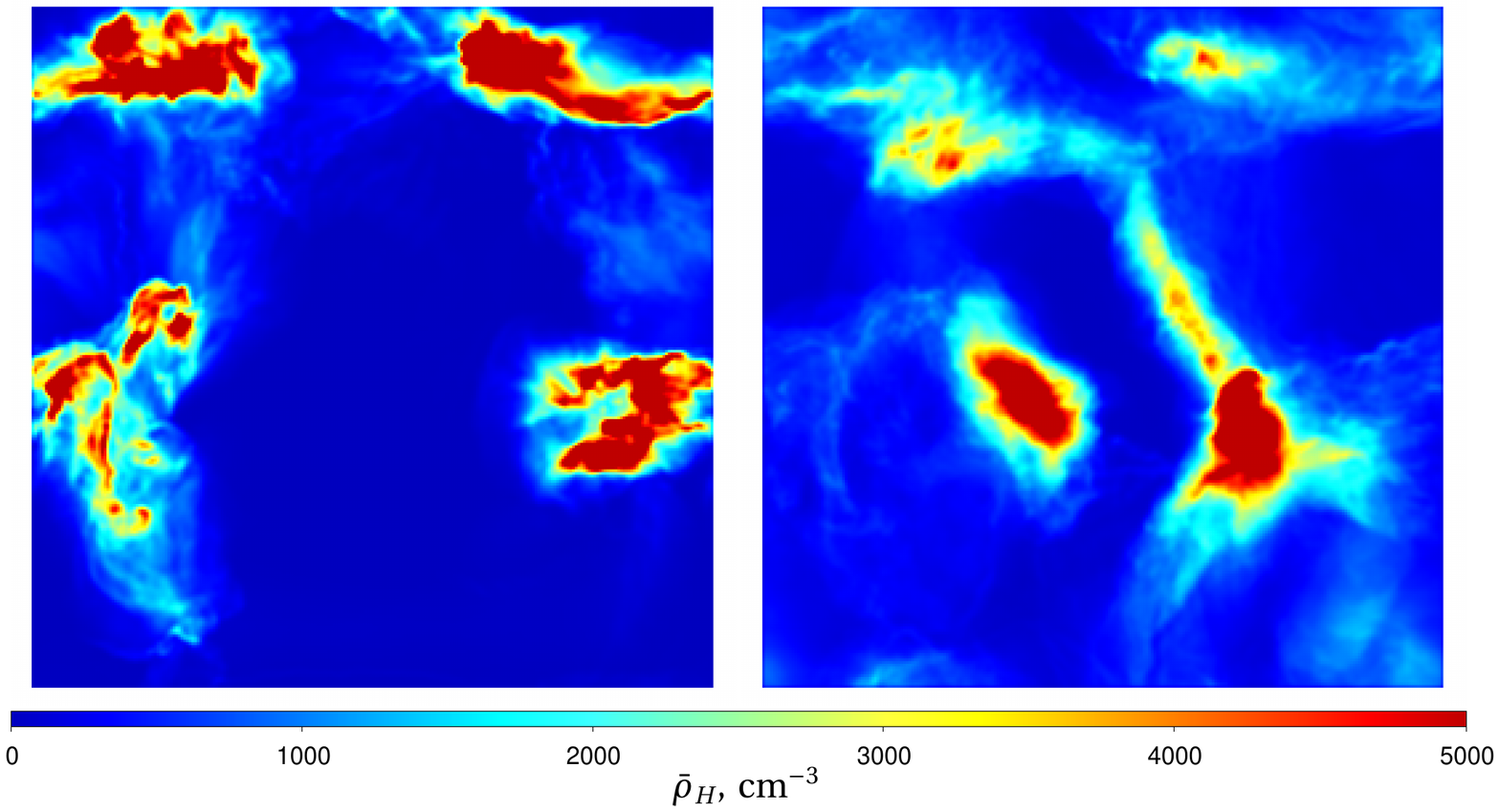}
\includegraphics[width=0.35\textwidth,viewport= 30 165 540 665]{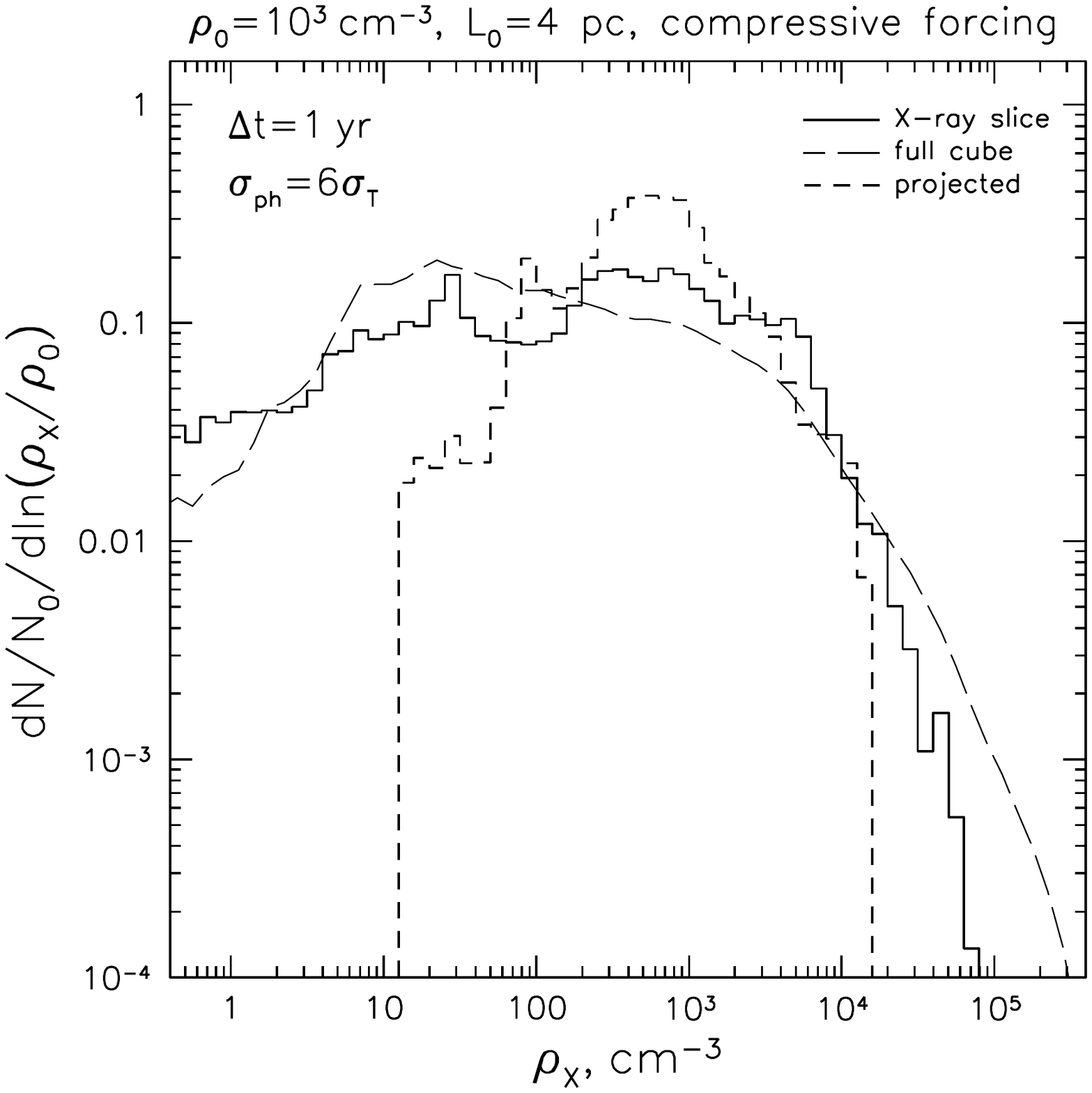}
\caption{Same as Fig.~\ref{f:proj_fed_sol_c3_s6_t1} but for compressively-forced turbulence.}
\label{f:proj_fed_com_c3_s6_t1}
\end{figure*}

Let us start considering statistical properties of X-ray reflection on a molecular cloud (or some part of it) represented by a box of simulated supersonic isothermal turbulence by \cite{2008ApJ...688L..79F}. As {these simulations are scale-free by design}, we could adjust the scales so that resulting configurations would resemble clouds of various size and, most importantly, opacity. We set the size of the box equal to $L_0=4$ pc and gas mean density equal to either $\rho_0=10^{3}$ or $10^{4}$ cm$^{-3}$. This results in the mean column density at level $ N_{H}\sim \rho_0 L_0\sim10^{22}$ cm$^{-2}$ and $\sim10^{23}$ cm$^{-2}$ for the first and the second case, respectively. {The corresponding masses are $M \sim m_{p}\rho_0 L_0^3\sim 2\times 10^{3}{\rm M}_{\odot}$ and $\sim 2\times 10^{4}{\rm M}_{\odot}$. }
{This choice of the scales results in rough consistency between the resulting physical properties of the simulation boxes and real molecular clouds for the main simulation parameter, namely a Mach number of $\sim5$ (and resulting velocity dispersion $\sim1$ km/s)
\citep[e.g.][]{1981MNRAS.194..809L,1987ApJ...319..730S,1992A&A...257..715F,2010ApJ...723..492R}.}

Given a 3D distribution of the molecular gas, X-ray reflection is modelled in the following (simplified) way: illumination proceeds in the plane-parallel manner along the $x$-axis of the cube from right to left, and {is then} observed by a distant observer located at $-\infty$ of the $z$-axis (as illustrated by Fig.~\ref{f:sketch}). Attenuation of the incident and outgoing X-ray radiation is applied with an effective cross-section $ \sigma_{\rm ph}=6\sigma_{\rm T}$ that {mimics photoelectric absorption by neutral gas} with solar metallicity in the 4-8 keV energy band. Having set the front propagation velocity $\varv_z=0.7c$, the thickness of the illuminated layer is then determined by {the} duration of the flare, while its age defines the position of this layer inside the box. In order to minimise distortions due to boundaries of the box, we will consider a situation when the illumination front passes through the middle of the box (see Fig.~\ref{f:sketch}).


{The X-ray} reflection signal is calculated according to Eq.\ref{eq:ix} for each simulation voxel that falls inside the illumination {region,} with the incident X-ray flux being correspondingly attenuated by {all the gas} located between the voxel and the primary source (located at $+\infty$ of the  $x$-axis, see Fig. \ref{f:sketch}). Further attenuation is applied due to the gas between the voxel and the observer (located at $-\infty$ of the  $z$-axis, see Fig. \ref{f:sketch}). After integration along each line-of-sight, one gets a surface brightness map of the X-ray reflection, which is then linearly converted into a corresponding average gas density map inside the layer. 

\subsubsection{Solenoidal vs. compressive forcing}
\label{sss:forcing}

{Left panels in Figures \ref{f:proj_fed_sol_c3_s6_t1} and \ref{f:proj_fed_com_c3_s6_t1} show the result of this procedure for boxes of solenoidally-forced (Fig. \ref{f:proj_fed_sol_c3_s6_t1}) and compressively-forced (Fig. \ref{f:proj_fed_com_c3_s6_t1}) turbulence with $\rho_0=10^{3}$ cm$^{-3}$ and flare duration $\Delta t=1$ yr. For comparison, the middle panels of these figures show maps} of {the average density found by volume-weighted integration along the line-of-sight across the whole box {(which is an equivalent of the column density map of the molecular gas inside the box)}}. Clearly, line-of-sight averaging smears the sharp substructures exemplified by the density map derived from X-ray reflection. Drastic difference in the dynamic range of these two maps is clearly visible.

This point can be even better visualised by comparing gas density \textit{PDF} that would be derived from the X-ray reflection and from a column density image. {Right panels in Figures \ref{f:proj_fed_sol_c3_s6_t1} and \ref{f:proj_fed_com_c3_s6_t1} demonstrate} that the gas density \textit{PDF} derived from X-ray reflection matches well the original gas density distribution of the whole box, even taken into account applied attenuation (important mostly in the high-density end of the distribution) and sampling variance due to relatively small volume of the illuminated layer. The characteristic dynamic range of these \textit{PDFs} is naturally much broader than for the integrated density map, which is quite narrowly centred on the mean gas density. As a result, the diagnostic power of the gas density \textit{PDF} recovered from X-rays should be significantly enhanced compared to the common column density studies.

\begin{figure}
\includegraphics[width=0.9\columnwidth, viewport= 30 180 550 680]{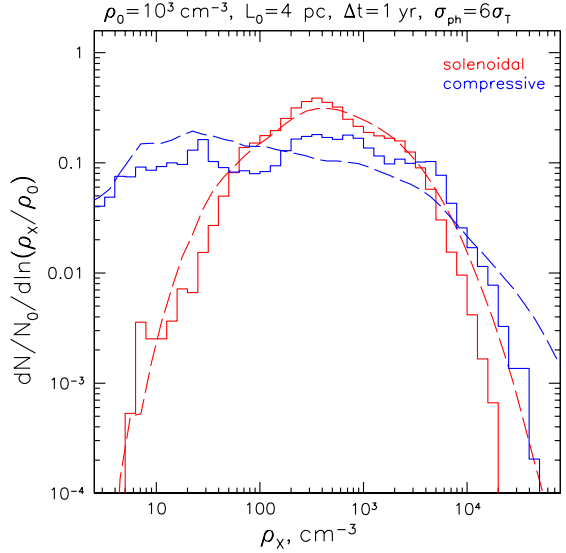}
\caption{Illustration of the possibility to distinguish solenoidal (red) vs. compressive forcing (blue). The solid {lines show the density \textit{PDF}s} of the slice with thickness corresponding to $\Delta t$=1 yr with effects of attenuation applied, the {long-dashed lines show} the original density $PDF$s of the whole cube.}
\label{f:pdf_s6_t1_n3}
\end{figure}

{Indeed, as is exemplified in {Fig.~\ref{f:pdf_s6_t1_n3}}, one can readily use the gas density \textit{PDF} recovered from X-rays surface brightness to distinguish solenoidal and compressive forcing of the turbulence with the same average Mach number. Supersonic turbulence with compressive forcing is characterised by $\approx\!3$ times larger width of {the} gas density \textit{PDF} compared to the case of solenoidal forcing \citep{2008ApJ...688L..79F}, and X-ray mapping in principle allows to observe corresponding differences both at the low-density and high-density ends of the distribution function {(see Fig.~\ref{f:pdf_s6_t1_n3}).}}

{However, the low-density end corresponds to low surface brightness of the reflected X-ray emission, so for real observations it might be affected by contaminating emission and statistical biases introduced by low-count statistics (see Section \ref{ss:observe}). These problems can be tackled by increasing {the} sensitivity of the available data and invoking additional information regarding sources of contamination, and we will discuss this more in Section \ref{ss:observe}.}

{On the other hand, the high-density end corresponds to compact dense substructures, which are embedded in bigger ones, resulting in a steady increase in the characteristic attenuation optical depth {$\tau$} for these regions (namely, in the case of the power-law scaling relation between density and size with the slope $\alpha=1$, {a} roughly logarithmic increase of {$\tau$ with density} might be expected). As a result, reflected X-ray emission from denser substructures should get suppressed compared to the optically thin prediction. Additionally, dense and compact substructures can "cast a shadow" on other regions that fall either behind them when seen from the primary source, or in front of them {when} seen from the observer.}

\subsubsection{Impact of attenuation}
\label{sss:attenuation}

\begin{figure}
\includegraphics[width=0.9\columnwidth, viewport= 30 180 550 680]{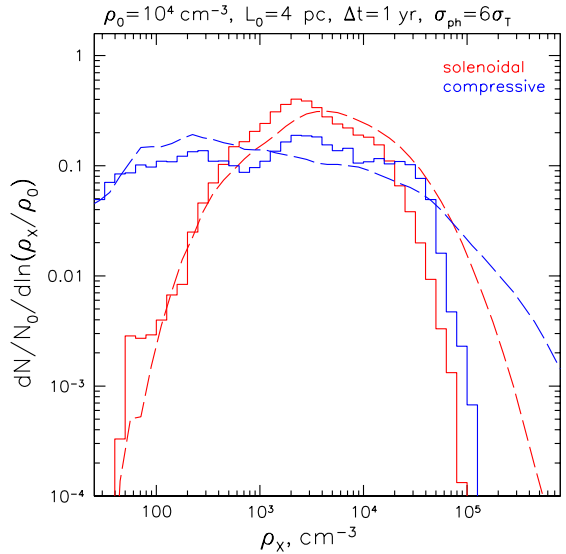}
\caption{
{
Same as Fig.~\ref{f:pdf_s6_t1_n3} but for a box with mean gas density $ \rho_0=10^4$ cm$^{-3}$. Notice prominent suppression of the high density tail due to strong attenuation of the densest regions of the box.
}
}
\label{f:pdf_s6_t1_n4}
\end{figure}

{
For the simulation boxes considered above and normalised to average density $ \rho_0=10^3$ cm$^{-3}$, the impact of attenuation effects on the shape of the density \textit{PDF}, however, stays  negligible down to the smallest scales (for the effective X-ray attenuation cross-section {in the 4-8~keV band} $\sigma_{\rm ph}\approx6\sigma_{\rm T}$, fiducial for neutral gas with solar metallicty). Taking the average density equal to $ \rho_0=10^4$ cm$^{-3}$ makes the suppression effect clearly distinguishable via {a steeper $PDF$ decline} at the high density end (see Fig.~\ref{f:pdf_s6_t1_n4}). Indeed, the column density at the largest scale is $\sim10^{23}$ cm$^{-2}$  and it increases by at least a factor of several for scales with $\rho\gtrsim 10\rho_0$, surpassing the threshold column density for the optically thin limit, viz. $N_{H}\simeq3\times10^{23}$ cm$^{-2}$. As a result, the reconstructed shape of the gas density \textit{PDF} at such densities is significantly distorted, both in terms of the high-end suppression and flattening at the pivot scale due to ``horizontal migration'' of the dense regions to lower reconstructed densities.  Clearly, for these high density regions the {information on the original shape of the true gas density \textit{PDF} is lost.}}

\subsubsection{Sampling variance}
\label{sss:variance}
\begin{figure}
\includegraphics[width=0.9\columnwidth, viewport= 30 180 550 680]{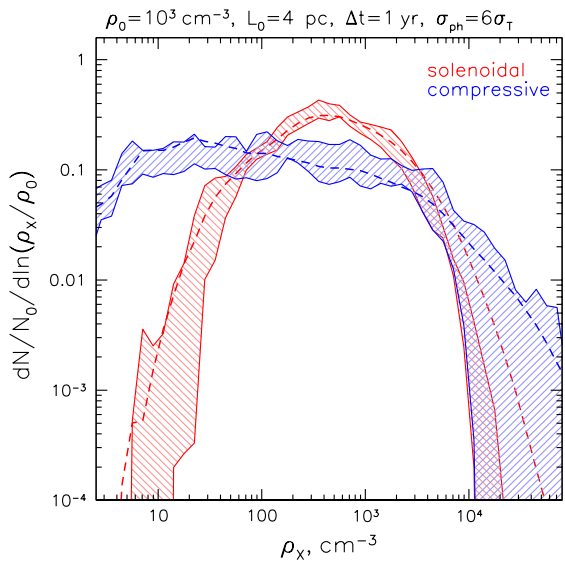}
\caption{
{{Increasing magnitude of the sampling variance at the high-density end} for the X-ray-recovered gas density $PDF$ in a box with solenoidal (red) and compressive (blue) supersonic turbulence. The boundaries of the hatched regions correspond to the {spread} of {the $PDF$} {values} constructed from individual 1-yr-thick slices, separated by 2 years each,{ the dashed lines show the original full-cube $PDF$s}.
}
}
\label{f:pdf_s6_t1_n3_var}
\end{figure}

{
Additional complication inherent to the high density end of the \textit{PDF} is associated with sampling variance, arising due to statistical variation in the number of the rarest dense substructures falling inside the thin slice of the illuminated gas. We illustrate this effect in Fig.~\ref{f:pdf_s6_t1_n3_var}, where the variance of the \textit{PDF} measured in 5 different slices (separated by approximately 0.5 pc) across the cloud is shown by the shaded regions.  Clearly, the combined impact of absorption and sampling variance hinders using of the high density end for turbulence diagnostic even for clouds with mean column density $\sim 10^{23}$ cm$^{-2}$.      
}

{
It is easy to see from Fig.~\ref{f:pdf_s6_t1_n3_var} that the reconstructed $PDF$ shapes for solenoidal and compressive turbulence are markedly different for densities below $\sim 5 \rho_{0}$, even taking into account absorption and sampling variance. Thus, one should focus on the $PDF$ shape approximately an order of magnitude around the mean density, where {{the} much more narrow distribution associated with} solenoidal forcing should clearly reveal itself. This is also particularly true for the low-density end, where turbulence with compressive forcing possesses prominent under-dense regions, or voids, constituting a very significant fraction of the cloud's volume while contributing very little to its total mass. In principle, the presence of such voids can be used as a unique diagnostic, which is accessible only in X-rays, since their detection in molecular species might be strongly hindered by effects of projection and changes in molecular excitation in such relatively low-density environments. {Of course, in the real molecular clouds, similar voids might appear as a result of the star-formation-related feedback (e.g. in the form of supernova explosions), so the practical application of this diagnostic is likely very problematic. }
}            

%
\subsection{Global ISM simulations}
\label{s:global}

\begin{figure*}
\includegraphics[width=1.\textwidth,viewport= 20 330 580 520]{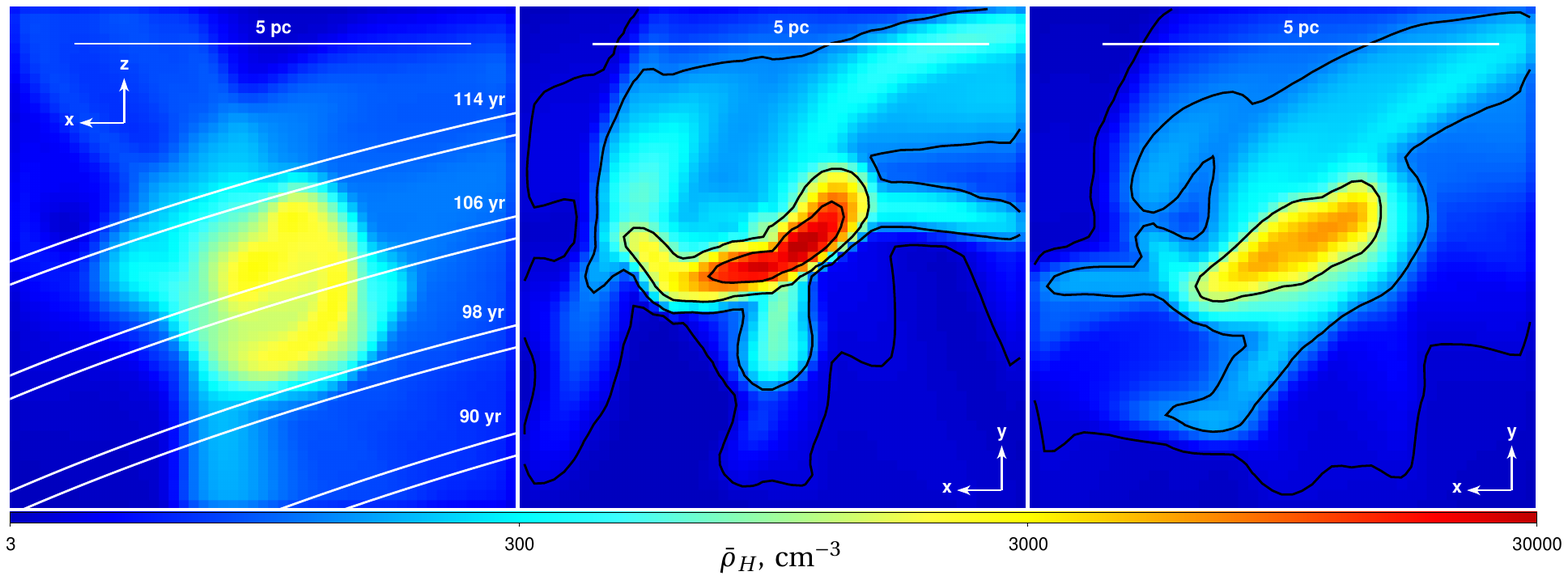}
\caption{Illumination of the (6pc)$^3$ cubic cut-out {from the SILCC-Zoom simulation for {molecular cloud MC2 \citep{2017MNRAS.472.4797S}} by a short flare. The geometry of} illumination is identical to the one illustrated in Fig.~\ref{f:sketch}. \textit{Left}. The average gas density map {(cm$^{-3}$)} as viewed from the top of {the box} along the $y$-axis. Locations of the illumination front at different times is depicted and labelled accordingly. \textit{Center}. X-ray reconstruction of the density {(cm$^{-3}$)} field in the picture (viz. $xy$) plane from the {slice at $t=106$ yrs}. \textit{Right}. {Map of the gas density {(cm$^{-3}$)} averaged {by volume-weighted integration} along the line of sight (viz. $z$-axis), which essentially corresponds to commonly observed column density map of the region.} The colour-coding is identical for the three panels and ranges from 3 cm$^{-3}$ to $3 \times 10^4$ cm$^{-3}$ on logarithmic scale.}
\label{f:topproj_mc2}
\end{figure*}

{
Real molecular clouds are far from being comprehensively represented by isolated boxes with isothermal turbulence, especially at the low-density end, where boundary conditions set by their parent environment must be taken into account, and at the high-density end, where the effects of self-gravity and stellar feedback cannot be neglected. }

{
Indeed, molecular clouds are not isolated objects, but rather strongly over-dense regions of the interstellar medium, and there exists a relatively smooth transition between the average low density ISM, $n\lesssim 1$ cm$^{-3}$, to denser envelopes of molecular clouds with $n\sim 10-100$ cm$^{-3}$, and finally to {the densest} regions constituting molecular clouds themselves,  $n\gtrsim 10^3$ cm$^{-3}$.}

We take advantage of the output from the SILCC-Zoom simulation for {molecular cloud MC2 \citep{2017MNRAS.472.4797S}}, focusing on a cubic (6 pc)$^3$ cut-out from it, centred on the cloud-like structure itself. The mean gas density in this cube is $\rho_{0}\approx 230$ cm$^{-3}$, reaching up to several $10^{6}$ cm$^{-3}$ in its central parts, and even more in embedded compact cores. 

{
{The geometry} of illumination is identical to the one we adopted for outputs of ideal supersonic turbulence simulations before. Namely, {the} $z$-axis is directed along the line of sight away from the observer, while {the $x-$ and $y-$axes} form {the plane of the sky} with {the} $x$ direction coinciding with the direction of the illumination (cf. Fig.~\ref{f:sketch}). The top view of {the} average gas density (i.e. integrated {with volume-weighting} over the $y$-axis) is shown in the left panel of Fig.~\ref{f:topproj_mc2}. }

{
The densest part of the cube, i.e. the molecular cloud itself, is located right in its central few parsecs, so it gets illuminated only during certain period of time, viz. for $t$ in between $\approx 100$ yrs and $\approx 115$ yrs for the illuminating source located at (-10 pc, 0 pc, -15 pc) with respect to the centre of the cube. Projections of the 1-yr-thick illumination front for $t$ ranging from 90 yrs to 114 yrs are depicted in the left panel of Fig.~\ref{f:topproj_mc2}. Such disposition (and corresponding age of the flare) is very similar to the one inferred for the currently brightest illuminated molecular complex in the Galactic Center region \citep[see e.g.][]{2017MNRAS.465...45C}.}

{
The central panel of Fig.~\ref{f:topproj_mc2} shows {the} X-ray reconstruction of the gas density for a slice illuminated at $t=106$ yrs, i.e. when the front {is} passing directly through the center of the cloud. For comparison, the right panel shows the gas density averaged {by volume-weighted integration} along the line of sight. Naturally, the density map reconstructed from a thin slice allows to reveal the densest compact central region with $\rho$ in excess of $10^{4}$ cm$^{-3}$ (cf. the black contours marking levels of 10, 100, $10^3$ and $10^4$ cm$^{-3}$ in the central and the right panels of Fig. \ref{f:topproj_mc2}).}

\begin{figure}
\includegraphics[width=0.9\columnwidth, viewport= 30 180 550 680]{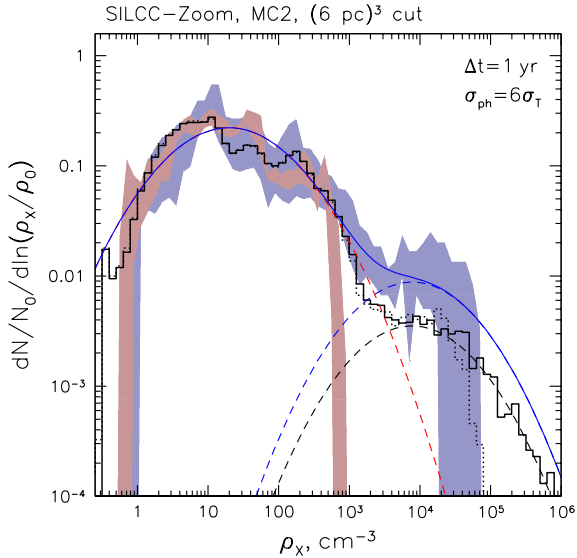}
\caption{
{Gas density \textit{PDF} reconstructed for the (6pc)$^3$ cubic cut-out {from the SILCC-Zoom simulation for {molecular cloud MC2 \citep{2017MNRAS.472.4797S}}}. The original full-cube $PDF$ is shown as black solid histogram, while the dotted histogram shows impact of attenuation (with the cross-section $\sigma_{\rm ph}=6\sigma_{\rm T}$) on the high-density end of this \textit{PDF}. Blue shaded region corresponds to $PDF$ reconstructed from six 1-yr-thick X-ray slices (separated by 2 years each) of the central densest region of the box, the red shaded region corresponds to the same but for a region lying in front of the central densest part (which is illuminated $\sim10$ yrs earlier). The dashed curves show log-normal approximations for the low-density and high-density parts of the \textit{PDF}. The solid blue line shows the combined model for the density $PDF$ of the central slices.
}
}
\label{f:pdf_mc1_s6}
\end{figure}

{
The gas density $PDF$ derived from six 1-yr-thick slices separated by 2 years with $t$ ranging from 100 to $110$ yrs is shown as blue shaded region in Fig.~\ref{f:pdf_mc1_s6}. The spread of this region {shows} the amplitude of the sampling variance, i.e. the difference between $PDF$s extracted from individual slices. The original gas density $PDF$ of the whole cube is shown as solid black histogram. 
}

{
Clearly, the gas density $PDF$ extracted from X-ray slices of the densest part of the box matches well the full-box $PDF$ for $\rho\lesssim 10^3$ cm$^{-3}$, but differs {increasingly} at higher densities. This difference, however, is {simply} a  result of the statistical averaging taking place in the case of the full-cube $PDF$. Indeed, the gas density $PDF$ extracted from {four slices} which lie in front of the densest central region of the box (corresponding to $t$ {from 90 to 96} yrs with 2-yr spacing, see Fig.~\ref{f:topproj_mc2}), demonstrates that the $\rho<10^3$ cm$^{-3}$ part of the density $PDF$ corresponds to the relatively lower density envelope of the cloud, while the dense part corresponds to the cloud itself.}

{
It can be seen that the overall density statistics derived from X-ray illuminated slices does reproduce the actual gas density $PDF$ very well {over a broad} range, spanning 3-5 orders of magnitude. Clearly, the low-density end, representing the envelope of the cloud, is robustly probed both with slices containing the cloud itself and those lying in front of it. Gas with such density constitutes the bulk of the volume and its $PDF$ shape can be vaguely described as log-normal {\citep[see also][]{2019MNRAS.482.5233K}} with mean $\rho_0\simeq100$ cm $^{-3}$ and $\sigma_s\simeq 1.7$ (see the red dashed line in Fig.~\ref{f:pdf_mc1_s6}). }

{
In the high-density part, one can see a dramatic change in the gas density $PDF$ as {the} X-ray illumination front enters the cloud. {Namely, an additional quasi-log-normal tail appears with the mean density $\rho\sim {\rm few}\times 10^4$ cm$^{-3}$ and $\sigma_s\simeq 1.7$ as well} (see the black dashed curve for the whole cube $PDF$ and the blue dashed curve for slices passing through the central region in Fig.\ref{f:pdf_mc1_s6}). The combined model constructed of the two log-normal distributions is shown in Fig.\ref{f:pdf_mc1_s6} as {the solid} blue line. Clearly, having $PDF$s measured over a time-span of $\gtrsim 15$ yrs, one can readily separate the gas density $PDF$ of the molecular cloud itself from the contribution of {the surrounding} ISM.}

{The gas densities around $\sim$few$\,10^3\,{\rm cm}^{-3}$, where the two log-normal parts overlap(see Fig.~\ref{f:pdf_mc1_s6}), correspond to the intermediate gas phase transitioning from the atomic to molecular state. The actual shape {of the density $PDF$ in this} region is sensitive to details of the gas thermal stability and the strength of the large-scale turbulent driving and dissipation \citep{2011ApJ...733...47W}. The X-ray reflection indeed allows this part of the $PDF$ to be well reconstructed and used for more advanced diagnostics.
}

{
An important difference between the full-cube $PDF$ and the one probed with X-ray reflection is obvious {at densities $\gtrsim 10^4$ cm$^{-3}$}, where X-ray attenuation suppresses {the} reflection signal from the densest parts. Indeed, although the intervening column density is relatively low, $\sim \rho_{0}L\lesssim 10^{22}$ cm$^{-2}$ for {the} bulk of the volume, it steadily increases for regions with high densities due to {the} hierarchical structure of the cloud, {such} that the regions with $\rho\gtrsim 10^{5}$ cm$^{-3}$ get fully attenuated (see the black dotted line in Fig.~ \ref{f:pdf_mc1_s6} {which shows the} impact of attenuation on the full-cube $PDF$). This effect, of course, strongly affects the $PDF$ of individual X-ray slices, in this case the exact {value of {the} density at which} the attenuation cut-off {takes place} is not so well defined due to strong impact of the sampling variance at the high-density end.
}

\begin{figure}
\includegraphics[width=0.9\columnwidth, viewport= 30 180 550 680]{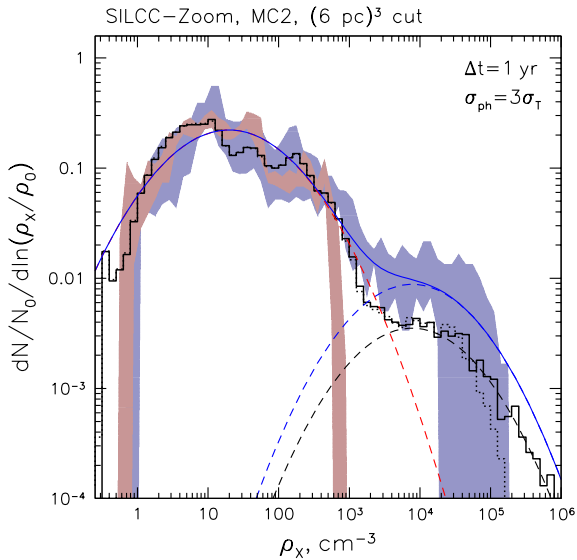}
\caption{
{Same as Fig.~\ref{f:pdf_mc1_s6} but with $\sigma_{\mathrm{ph}}=3\sigma_{\rm T}$. Notice substantial extension of the probed density range in the high-density tail.
}
}
\label{f:pdf_mc1_s3}
\end{figure}

{In principle, {the impact of the }attenuation can be somewhat weakened by selecting another spectral energy band where the effective attenuation cross-section is smaller than in the fiducial 4-8 keV energy band. The \textit{PDF} that can be reconstructed with $\sigma_{\mathrm{ph}}\approx3\sigma_{\rm T}$ is shown in Fig.~ \ref{f:pdf_mc1_s3}. One might expect such effective cross-section for instance for a narrower band from 6 to 7 keV (which still contains the Fe I fluorescent line, the most distinctive feature of the X-ray reflection) or at energies $\gtrsim10$ keV.}

{
Unfortunately, in practice one has to deal with a number of observational limitations, set by accessible sensitivity, angular and spectral resolution of the data.  Currently, these limitations prohibit using a narrow spectral window or harder X-ray range, but the next generation of X-ray observatories will allow such analysis to be performed. In the next Section, we discuss all the relevant observation-related issues in full detail.}


\section{Discussion}
\label{s:discussion}

{
Above we considered \textit{inevitable} distortions inherent to X-ray probing of the gas density statistics arising due to finite duration of the flare and obscuration of the densest regions even for highly penetrative X-ray emission. 
{The} real data are always prone to a certain level of noise and information loss, which potentially lead to biases in the measured quantities. 
}{
Another important complication is that one always has to {deal with} contamination of the detected signal by X-ray emissions of different nature. This is particularly the case for the very crowded region of the Galactic Center, hosting both {an} abundant population of point sources and plenty of extended structures \citep[e.g.][]{2004ApJ...613..326M,2009ApJS..181..110M,2013MNRAS.428.3462H,2013MNRAS.434.1339H,2015MNRAS.453..172P}.}

{In order to minimise {the} possible impact of all these issues one has to ensure that the quality of the data is sufficiently high to allow robust measurements to be performed. Next we thoroughly discuss all these complications and the ways to handle them given {the currently available and the foreseen quality of data.}}

\subsection{Implications of limited sensitivity and angular resolution}
\label{ss:observe}

\begin{figure}
\includegraphics[width=\columnwidth,viewport=30 200 600 670]{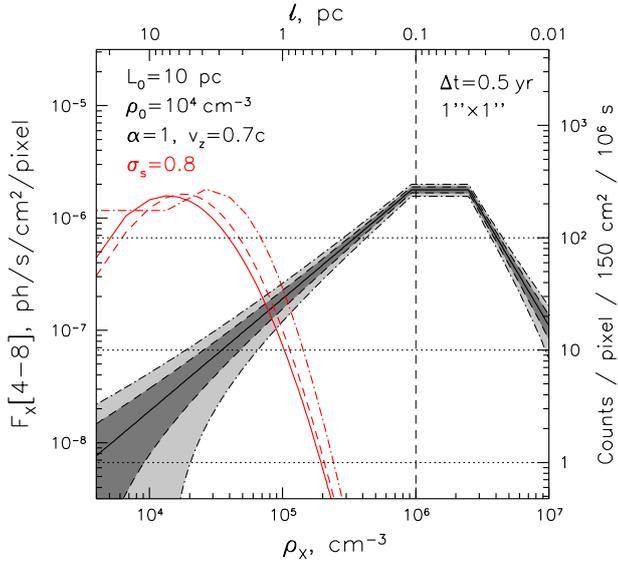}
\caption{{
{Similar to Fig.~\ref{f:tf1} but showing the relation between gas density and photon flux per 1''x1'' pixel for the reflected emission of a 0.5-yr long flare { taking into account} the impact of finite angular resolution. Smearing with {the instruments'} resolution leads to {a strong} suppression of the density-surface brightness relation at the high density end, introducing the second break in this relation. The right {axis shows the} corresponding expected number of 4-8 keV counts to be detected by an instrument with 150 cm$^2$ of effective area in this band after a $10^6$ { s exposure}. 
The shaded regions show the expected Poisson uncertainty in the surface brightness measurement after such (dark grey, dashed boundary) and 4 times shorter (light grey, dash-dotted boundary) observation. The corresponding Eddington bias distortions of the original log-normal distribution (solid red curve, $\rho_0=10^4$ cm$^{-3}$, $\sigma_s=0.8$) are shown as dash-dotted and dashed red curves, respectively, { in a way identical to Figure \ref{f:tf1}.}}
}
\label{f:binedd}
}
\end{figure}

{
Angular resolution of modern X-ray telescopes {does not} allow resolving structures {smaller than} $\sim 1$ arcsec, viz. $0.04$ pc at the distance to the Galactic Center, with \textit{Chandra} and $\sim 5$ arcsec, viz. $0.2$ pc, with \textit{XMM-Newton}. As a result, the X-ray flux produced by the substructure of size $l$ is smeared over {a} region of {the} Point Spread Function (PSF) size, $\delta x$. If $l<\delta x$, the resulting surface brightness gets suppressed by {a factor of} $(\delta x/l)^2$. Once again, {assuming a} power-law relation between  $\rho_{H,l}=\rho_{H}(l)\propto l^{-\alpha}$, {one can expect a} break in the $I_{X}(\rho_{H})$ relation due to $\rho_{H}(l)^{-2/\alpha}$ suppression at densities corresponding to $l<\delta x$. This effect is illustrated in Fig. \ref{f:binedd} {(for the case of the powerlaw size-density scaling relation with $\alpha=1$ and $\rho_0=10^4$ cm$^{-3}$ at $L_0=10$ pc).}
}

{
Clearly, {as long as} the characteristic scale {which is} set by {the} duration of the flare, $\delta z=v_{z}\Delta t$, is larger than the scale  {set} by {the} angular resolution of the data, this suppression is not of primary importance. For {the brightest clouds in the vicinity of Sgr~A*} and {the} angular resolution available with \textit{Chandra}, this is the case for $\Delta t\gtrsim0.15$ yr. For  \textit{XMM-Newton}, however, this suppression becomes dominant already for $\Delta t\sim0.75$ yr, effectively prohibiting {unambiguous} probing of structures below few$\times$0.1 pc with this instrument. 
}

{
In {practice}, the effective angular resolution of the data, especially for probing the low surface {brightness} end of the distribution function, is set by the available sensitivity. Indeed, for some fiducial values of the cloud's, flare's and observation parameters, Eq.\ref{eq:ix} gives }
\begin{eqnarray}
\label{eq:cx}
C_X=0.2~\mathrm{cts~pixel^{-1}}~\left(\frac{L_X\Delta t}{3\times10^{46}~\mathrm{erg~s}}\right)\left(\frac{\varv_z}{0.7c}\right)\left(\frac{D_{sc}}{30~\mathrm{pc}}\right)^{-2}\times\nonumber\\
\times\left(\frac{\bar{\rho}_H}{10^{4}\mathrm{cm^{-3}}}\right)\left(\frac{dx}{\mathrm{1~arcsec}}\right)^2 \left(\frac{A_{\rm eff}}{\mathrm{150~cm^2}}\right)\left(\frac{t_{\rm exp}}{\mathrm{10^5~s}}\right),~~~~~~
\end{eqnarray}
where $dx$ is the data pixel size, $A_{\rm eff}$ is the characteristic effective area of the telescope in 4-8 keV band, and $t_{\rm exp}$ is the observation's exposure time. The selected fiducial values for these parameters match well the parameters of the currently available data provided by $Chandra$ \citep[e.g.][]{2017MNRAS.471.3293C}.

{ The low count number statistics results in {a} high uncertainty of the measured fluxes and hence strong statistical deviations from the linear $I_{X}-\rho$ relation. This gives rise to the so-called Eddington bias, a characteristic distortion of the measured distribution function due to 'horizontal migration' of individual data points. {A similar} situation takes place for instance for steeply declining ends of luminosity functions, where upward statistical fluctuations of fainter sources can easily dominate over {the} number of {much rarer} intrinsically bright sources.}

{
The problem {at hand} can be approximately treated as the case of {a} log-normal parent distribution with flux measurement {subject to} Poisson noise of the photon counting statistics. The resulting distorted distribution function {can then} be formulated in integral form and evaluated numerically, as has been done in numerous papers in bio-statistics, where Poisson-lognormal convolution is relevant on many occasions \citep[e.g.][]{Bulmer1974}. In fact, numerical integration can be avoided, since the approximation provided by the saddle point integration turns out to be sufficient in many cases \citep{Izsak2008}. We briefly describe this method in Appendix \ref{s:appb}, so it can be easily implemented in practice in future studies.} {The resulting distortion of the measured $PDF$ function is illustrated in Fig. \ref{f:binedd}. One can see that for $t_{\rm exp}\sim10^6$ s (i.e. 1 Ms) the overall amplitude of the $PDF$ distortion is small, but for $t_{\rm exp}$ $\sim4-5$ times smaller (i.e. 200-250 ks, which is comparable to the deepest currently available data), {the} $PDF$ measurement is strongly affected {for the major part of the volume}. {Next we} discuss the implications of this distortion for {the} estimation of parameters characterising the $PDF$ shape.}

{
It is worth mentioning here that the uncertainty of {the} surface brightness measurement {can be} higher than purely Poisson uncertainty {if spectral} separation of the reflected X-ray emission from the contaminating background and foreground emission {is needed}. In the case of (apparently) diffuse contaminating emission, the linear component separation method can be used that takes into account different spectral shapes of various components. Robust operation of this method typically requires a few tens of total counts to be detected in 4-8 keV band in order to properly sample relevant spectral features of the characteristic spectral models (see Appendix \ref{s:appa}).}

{Thanks to temporally and spatially smooth (and in principle predictable) behaviour of the most important contaminating diffuse component, one can significantly diminish biases of {the} component separation technique even for {a very} low number of counts in each individual pixel. {For instance, the spatial distribution of mid-infrared light is believed to trace the distribution of the old stellar population, allowing one to predict the intensity of the corresponding Galactic X-ray Ridge emission with accuracy of few 10\% \citep[e.g.][]{2009Natur.458.1142R}}. {Our tests show that the overall boost of uncertainty in the measured reflection signal intensity typically amounts to a factor of several. Possible impact of additional sources of contamination, namely point sources, second scatterings and X-ray reflection due to intervening structures along the line-of-sight in {a} multi-flare scenario, are thoroughly discussed in Appendix \ref{s:appa}.  }}

\subsection{Accuracy of \textit{PDF} reconstruction  with current and future data}
\label{ss:accuracy}
\begin{figure}
\includegraphics[width=\columnwidth,viewport=30 200 550 690]{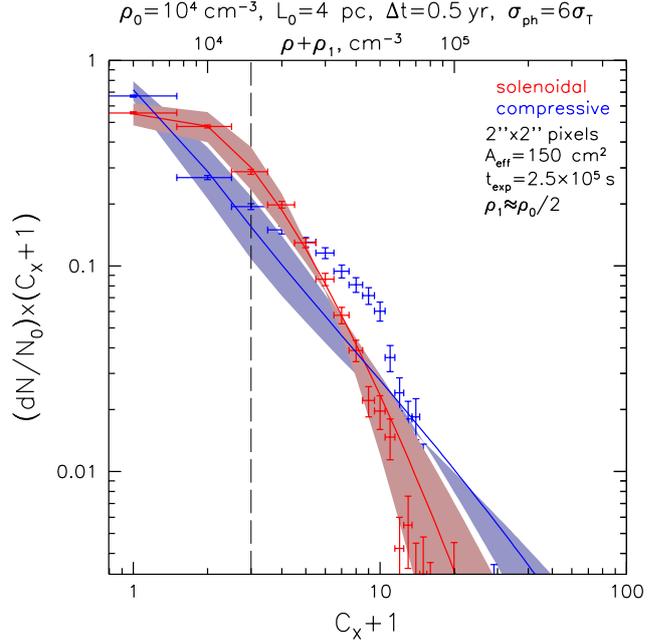}
\caption{
{Simulated distributions of the detected number of counts per $2\arcsec\times2\arcsec$ pixel (4-8 keV band, $C_{\rm X}$+1 is used for the $x$-axis instead of $C_{\rm X}$ for better visibility on logarithmic scale) after a $250$ ks-long observation of the reflected emission from boxes of isothermal supersonic turbulence with size $L_0=4$ pc and mean density $\rho_0=10^4$ cm$^{-3}$ (red points for solenoidal forcing, blue points for compressive forcing, see Section \ref{s:simulations}) with a $Chandra$-like instrument ($A_{\rm eff}\sim$ 150 cm$^2$ at 4-8 keV). Parameters of the flare ($\Delta t\sim0.5$ yrs and $L_{X,4-8}\Delta t\sim3\times10^{46}$ erg) and cloud's disposition ($D_{sc}\approx30$ pc, $v_{z}=0.7$c) are similar to the ones inferred for the {Sgr A} molecular complex which is currently the brightest one in reflected emission \citep{2017MNRAS.471.3293C}. Horizontal 'error-bars' of the points equal 1 (i.e. simply depicting size of the bin), while vertical error-bars correspond to 1-$\sigma$ variance due to shot noise in number of pixels per each $C_{\rm X}$ bin. Log-normal approximations (taking into account Eddington bias at the low-surface brightness end) with expected values of the width ($\sigma_{s}$=1.15 and $\sigma_{s}$=1.8 for solenoidal and compressive forcing, respectively) are shown with solid lines, while the shaded regions correspond to $\pm20\%$ variation in $\sigma_s$ (with fixed average density and overall normalisation).{ The top axis shows the corresponding gas number density with addition of the reference density $\rho_1=5\times10^{3}$ cm$^{-3}$ giving 1 count per pixel over full exposure time. Vertical dashed line marks the position of the mean density.}
}  
}
\label{f:pdfobs2e5}
\end{figure}

\begin{figure}
\includegraphics[width=\columnwidth,viewport=30 200 550 690]{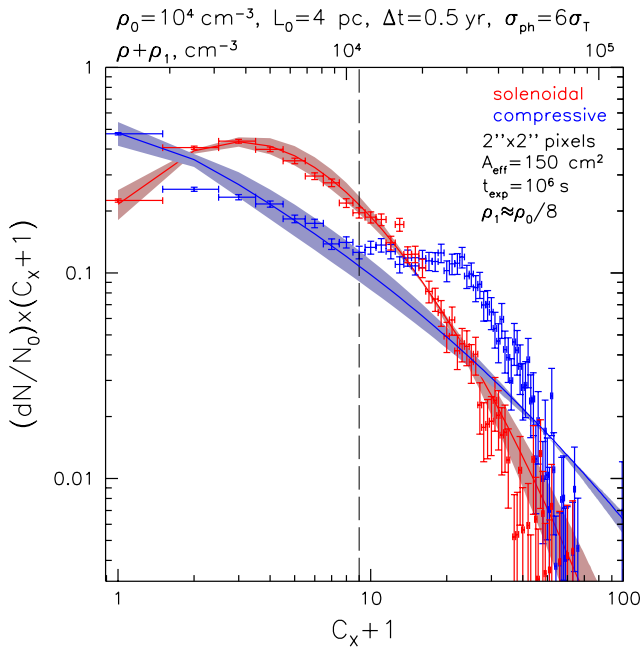}
\caption{Same as Fig.\ref{f:pdfobs2e5} but for a 1Ms-long observation. The shaded regions correspond now to $\pm10\%$ variation in $\sigma_s$. { The reference gas density $\rho_1$ equals $1.25\times10^{3}$ cm$^{-3}$} in this case.}
\label{f:pdfobs1e6}
\end{figure}

{
Having in mind the limitations and distortions inherent to {an} X-ray measurement of the gas density $PDF$ described above, we can quantify the accuracy of the $PDF$ parameter estimation accessible with such data. Since it is of primary importance to probe the gas density $PDF$ down to $\sim0.1$ {pc} scales, {the} angular resolution of the data needs to be kept $\lesssim 2$ arcsec. The question is then how large is the dynamic range of scales that is accessible to robust X-ray probing with such binning of the data. Clearly, the largest scale corresponds to the size of the cloud, so the maximal dynamic range {spans 1-2 orders of magnitude} for the cloud's size ranging from 1 to 10 pc. This also means that the whole cloud typically contains $N_0\sim$few$\times10^2$ to few$\times10^4$ 2$\arcsec$x2$\arcsec$ pixels. }

{The surface brightness of the reflected emission at the largest scales is plausibly too low for robust measurement in individual pixels. Nonetheless, the \textit{mean} gas density in the cloud can be robustly inferred from the total measured reflection flux, given that this quantity characterises the cloud as a whole, so averaging across the biggest available scale does not introduce any bias. As a result, one can then readily calculate the expected number of counts $C_0$ to be detected for a pixel containing gas of the mean density with very high accuracy\footnote{This does not mean that the \textit{absolute} value of mean density can be measured with such high accuracy due to a number of uncertainties inherent to the X-ray probing technique. Although being of no direct importance for the $PDF$ \textit{shape} reconstruction, such absolute measurement would be of great value by itself, especially taking into account very weak dependence of X-ray reflection on chemical and thermodynamical properties of the gas.}.}

{ Reliable reconstruction of the $PDF$ shape is possible only if sufficient number of pixels with detected number of counts $C_{th}\gtrsim10$, i.e. at densities $\rho_{X}\gtrsim C_{\rm th}/C_{0}\rho_{0}$, are present.} {The} fraction of such pixels is $\sim Q(x)=0.5~{\rm erfc}\left(\frac{\ln(C_{\rm th}/C_{0})}{\sqrt{2}\sigma_s}\right)$ in the case of {a} log-normal distribution with dispersion $\sigma_s$, $x=\ln(\rho_{X}/\rho_{0})/\sigma_s$. This fraction equals $10^{-2}$ for $x=2.3$ (and $10^{-3}$ for $x=3.$), so that for {the} number of such pixels to be sufficient for $PDF$ reconstruction one needs {to have} at least $C_0\gtrsim \frac{C{\rm th}}{\exp(2.3\sigma_s)}\approx 1$ for $C_{\rm th}=10$ and $\sigma_{s}=1$. As can be seen from Eq.\ref{eq:cx}, { this means exposures of at least as few hundred kiloseconds with a $Chandra$-like instrument ($A_{\rm eff}\sim$ 150 cm$^2$ at 4-8 keV) are needed for reconstruction of a log-normal $PDF$ with 2$\arcsec$ spatial binning.}

{The $PDF$ shape is typically more complicated than purely log-normal even in the case of isothermal supersonic turbulence, especially in the high-density part where sampling variance, effects of opacity and angular smearing all might become pronounced. Due to this, it is very valuable not only to measure {the} characteristic width of the distribution in the form of $\sigma_s$ but to reconstruct its actual shape to be able to correct it for {the} possible impact of these effects. }

{We took advantage of the results presented in Section \ref{s:results} for the boxes of supersonic isothermal turbulence to check this possibility. Namely, we performed Monte Carlo simulations of observed distribution functions taking into account discrete sampling of the original $PDF$ function with superimposed Poisson noise in {the} measured number of counts. Naturally, such simulations allow us to reproduce {the} Eddington bias at the low-density end and the shot noise at the high-density end in a fully self-consistent manner.}

{Results of such simulations for a $Chandra$-like instrument ($A_{\rm eff}\sim$ 150 cm$^2$ at 4-8 keV), 2$\arcsec$ spatial binning, and 250 ks and 1 Ms exposure times are shown in Figs.~\ref{f:pdfobs2e5}~and~\ref{f:pdfobs1e6}, respectively. While the former roughly corresponds to the quality of the best currently available data \citep{2017MNRAS.471.3293C}, the latter demonstrates what might be achieved with a deep $Chandra$ exposure or with a very modest exposure with X-ray observatories of the next generation, i.e. $ATHENA$ and $Lynx$ \citep[see][for a dedicated discussion]{2019arXiv190306429C}. }

{ It useful to define a reference density $\rho_1$ corresponding to the expectation value of 1 count per pixel to be detected over the whole exposure time. From the Eq.\ref{eq:cx}, one can see that $\rho_1\sim4\times10^3$ and  $\rho_1\sim1\times10^3$ cm$^{-3}$ for t$_{\rm exp}$=250 ks and 1 Ms respectively. Only for densities a factor of 10 larger than this are not strongly affected by the statistical noise in the data.}

{ As can be seen from Fig. \ref{f:pdfobs2e5}, currently available data ($t_{\rm exp}\sim$ few 100 ks) { allows for} robust probing only for the high density part of the $PDF$, i.e. at $\rho_{x}\gtrsim3\rho_0$. However, including theoretical prescription for the Poisson-lognormal distortion (described in Appendix \ref{s:appb}) results in $\sim 20\%$ accuracy of {the} $\sigma_s$ measurement (see shaded areas in Fig. \ref{f:pdfobs2e5}). Since expected values of $\sigma_s$ are $\approx 1.15$ for solenoidal forcing and $\approx$1.8 for compressive forcing (given that $\mathcal{M}\sim5$ in both cases, cf. Eq.\ref{eq:sigmas}), the current data \citep[cf. the analysis performed in][]{2017MNRAS.471.3293C} provide only marginal possibility to distinguish between these two scenarios. Moreover, the additional noise introduced by necessity of spectral filtering of the contaminating emission should make the purely Poisson-lognormal model not applicable, especially at the low surface brightness end of the distribution.}

{The situation changes significantly in the case of 4-times longer (i.e. $\sim 1$ Ms-level) exposure (see Fig. \ref{f:pdfobs1e6}). Indeed, here the bulk of the volume is expected to produce {a} non-zero number of counts, and the {overall} $PDF$ shape can be well reconstructed over the dynamic range of a few tens. {The resulting uncertainty of the} $\sigma_s$ measurement would not exceed $10\%$, allowing firm conclusions on the dominant forcing mechanism to be done.}    

{As has been shown in Section \ref{ss:global}, the high-density part of the $PDF$ extracted from the realistic zoom-in simulations of the molecular clouds can be well approximated as log-normal with $\sigma_s\approx1.7$. Clearly, $\sim10\%$ accuracy {in measuring} $\sigma_s$ should be feasible with such data as well, given that opacity and flare's duration do not prohibit probing the scales down to $\sim$0.2 pc. {As mentioned earlier, at these scales a power-law tail might appear in the density $PDF$ as a result of self-gravity taking over the turbulent motions. We don't expect that this tail will affect parameter estimation for the bulk $PDF$, but it is of great interest to study this transition by itself, of course. }}  

{{ Once again, contamination of the reflected signal by diffuse emission of other nature as well as foreground and background point sources needs to be fully taken into account in the analysis of the real data. A  discussion of the resulting complications and the ways to overcome them is given in Appendix \ref{s:appa}.} In general, some of these issues might be treated by increasing the lower threshold for {the} minimum number of the detected counts per pixel in addition to exploiting information on temporal variations and spatial distribution of the contaminating signals. Fortunately, X-ray observatories of the next generation will benefit not only from $\sim10$-times higher sensitivity, but also from excellent spectral resolution ($\sim$few eV at 6.4 keV in the case of cryogenic bolometers) making possible more robust component separation (see Appendix \ref{s:appa}) and usage of narrower spectral bands (e.g. centred on the iron fluorescent line at 6.4 keV). }

{{An even} more ambitious opportunity would be to use {a} harder X-ray energy band, namely above 10 keV, where the impact of opacity is minimised. Indeed, $\sigma_{\rm ph}$ falls below $2\sigma_{\rm T}$ for $E>15$ keV, allowing to probe a factor of several higher densities. An additional advantage of this band is also decreased sensitivity of both X-ray albedo and attenuation on the gas metallicity. The most important limitation is, of course, set by angular resolution of the instruments operating in this energy band. Since component separation based on spectral decomposition is unlikely to be feasible in {the} hard X-ray band alone, a promising approach would be to combine the information obtained in different bands in order to single out effects of opacity (assuming uniform metallicity across the cloud).
}

\subsection{Probing {the} gas velocity field}
\label{ss:velocity}
\begin{figure*}
\includegraphics[width=0.83\textwidth,viewport= 30 265 560 570]{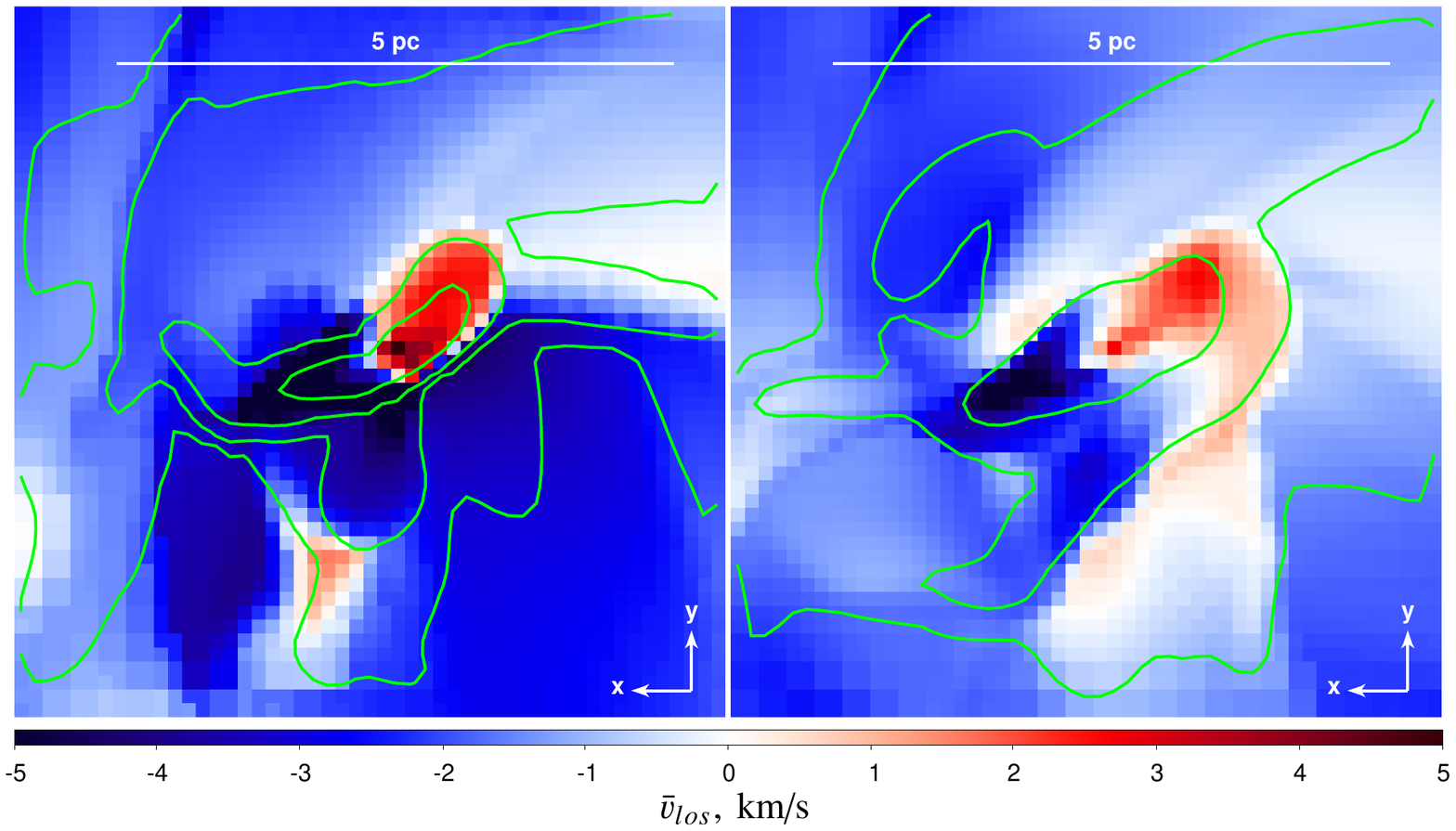}
\caption{
{
Line-of-sight velocity field (km/s) for an X-ray illuminated central slice (left) and full projected (6pc)$^3$ cubic cut-out (right) from the molecular cloud MC2 of the SILCC-Zoom simulation. The green contours correspond to the same levels of the average gas density as in the central and right panels of Fig.~\ref{f:topproj_mc2}. As the reflection front passes through the cloud, a number of such snapshots can be obtained allowing one to estimate also gradient of this velocity component along the line-of-sight.
}
}
\label{f:velx_mc1}
\end{figure*}

{
Above we considered intensity mapping of the reflected X-ray emission as a way to probe statistics of the gas density field. {The} next generation of X-ray observatories featuring both high collecting area and exquisite spectral resolution, $FWHM\sim$few eV at 6.4 keV, will allow a natural step forward to be made, namely spectral mapping, and, {in particular, measurement} of the fluorescent line centroid, width and {amplitude} of the line's ``Compton shoulder'' \citep[see a dedicated discussion of the prospects of future X-ray observatories in][]{2019arXiv190306429C}.
}

{In fact, the fluorescent line is a close doublet of the $K_{\alpha,1}$ and $K_{\alpha,2}$ components, separated by $\sim12$~eV (as can be seen in Fig.~\ref{f:shoulder}). The estimated natural width of each component is \citep[$FWHM\approx1.6$ eV, e.g.][]{1979JPCRD...8..329K}. However, the actual shape of the $K_\alpha$ complex is more than a combination of two Lorenzians  \citep[see, e.g.][]{2004A&A...414..377M,2016PhRvA..94d2506I}. This complexity may preclude accurate line broadening diagnostic for cases when Doppler broadening is intrinsically small. The line shift can also be affected, albeit to a lesser degree, and one might hope measuring the centroid energy with $\sim FWHM/10\sim$ few 0.1 eV accuracy at least for the brightest clumps in reflected emission.}

{
{Typical turbulent velocities are not very high, though: for gas with temperature $T\lesssim 100$ K, {the} sound speed amounts to $c_{s}=\sqrt{\gamma T/\mu m_{p}}\lesssim 0.6$ km/s (for $\gamma=1$ and the mean molecular weight $\mu=2.3$), so one might expect turbulent velocities $\varv \sim 3-6$ km/s for Mach number ranging from 5 to 10. The resulting variations in the centroid energy of the iron fluorescent line are $\sim$0.07-0.15 eV on the largest scales (assuming that {the} turbulent cascade is forced at scales comparable to the size of the cloud).} Since the line broadening is associated with velocity dispersion (along the line-of-sight) at scales comparable to the width of the illumination front, it is expected to be even smaller and hardly measurable using X-ray reflection.
}

{
Actually, superimposed shearing distortions of the velocity field might reach higher values, especially in the very extreme and dynamic environment of the Central Molecular Zone. Such a situation apparently takes place for the massive and compact Brick cloud \citep{2016ApJ...832..143F}. The observed gradient of its projected velocity field amounts to few tens km/s that can in principle be mapped with {the} next generation of X-ray observatories \citep[e.g. $ATHENA$ and $Lynx$, see corresponding discussion and an expected map of the fluorescent line's centroid offset for the Brick cloud in][]{2019arXiv190306429C}.
}

{This picture is very well reproduced using the outputs of global ISM simulations we discussed earlier. Fig.~\ref{f:velx_mc1} shows the projected velocity field that might be inferred from a slice of X-ray illuminated gas (left panel) and from the full central (6 pc)$^3$ cut-out of the SILCC-Zoom MC2 simulation box centred on the molecular cloud (see Section \ref{ss:global} for the discussion of the corresponding density fields). One can see a $\sim 10$ km/s velocity gradient in the central part ($\sim2$ pc) of the box, which is partially smeared out by line-of-sight averaging in the case of the full-cube projection. Naturally, this part of the volume is characterised by the highest gas density, and hence it might be expected to be bright enough in reflected emission for the intricate line centroid diagnostics. In addition to this, measuring variation of the line centroid with time, i.e. as the illumination front passes through the cloud, one can also estimate the line-of-sight gradient of the velocity.  This would allow getting an idea about the full 3D gradient in the line-of-sight component of the velocity field and possibly shedding more light on the primary driving mechanism.}

 {Another complication in the line centroid measurement might arise due to partial ionisation of iron in the probed molecular gas (this could be particularly relevant for clouds in the CMZ). For neutral iron, energies of the lines are known with high precision. For weakly-ionised iron (Fe$^+$-Fe$^{+9}$), calculations of { \citet[][]{2003A&A...410..359P} and \citet{2004A&A...414..377M}} suggest {a} very small change of the line centroid energy, consistent with the accuracy of their calculation of $\sim 7\,{\rm eV}$ at 6.4 keV \citep[see also earlier calculations of][which predict {a} larger wavelength shift]{1993A&AS...97..443K}. However, this uncertainty  translates into a velocity shift of $\sim$ 300 km s$^{-1}$, implying that {a} higher level of accuracy will be needed to keep {the} possible impact of partial ionisation under control.}
 
 
{A promising opportunity would be to use much higher spectral resolution accessible with radio observations of line emission from various molecular species by cross-correlating these data with the X-ray data collected over {a} time-span comparable to the light-crossing-time of the whole cloud. Clearly, this requires an extensive campaign of regular (separated by approximately duration of the flare) sufficiently deep X-ray observations to be performed over $\sim10$ yrs. Given that the Galactic Center region harbours our Galaxy's SMBH Sgr~A* and {a} rich variety of point sources and thermal and non-thermal diffuse emission \citep[e.g.][]{2015MNRAS.453..172P,2018PASJ...70R...1K}, it is very likely to be extensively observed in {the} future as well, providing the necessary X-ray coverage for one of the illuminated clouds as a by-product.   
}

{Another possibility is related with establishing association between {the} brightest X-ray clumps and sources of maser emission, plenty of which are detected in the Galactic Centre region \citep[e.g.][]{2016ApJS..227...10C}. Observations of masers can provide not only high precision line-of-sight velocity measurement, but also measurement of the proper motion in the plane of sky. In this synergy, X-ray density reconstruction would allow accurate location and environment characterisation of the maser region, while the maser itself would allow inferring the 3D velocity vector. A systematic campaign of sensitive regularly-spaced X-ray observations is again a key requirement for such kind of studies.}

\newpage
%
\section{Conclusions}
\label{s:conclusions}
   
{~~~~~Taking advantage of the numerical simulations aimed at reproducing {the inner structure of molecular clouds} (both in the framework of ideal isothermal supersonic turbulence and in zoom-in extractions from global ISM simulations), we have demonstrated that reflected emission of a short X-ray flare is indeed a powerful tool for recovering {the} statistics of the gas density field. The dynamic range of scales accessible for this technique is indeed sufficient to judge on characteristics of the underlying generic physical processes, e.g. the type of turbulence forcing.} 

{For nearly-lognormal gas density $PDF$s with a typical value of the width $\sigma_s\sim1-2$, {the statistical uncertainty in the measurement} amounts to { at least $\sim20\%$ with the currently available data if only Poisson noise is included.} Taking into account systematic uncertainties connected with the necessity of filtering contaminating diffuse signals, solenoidally- and compressively-driven turbulence can be distinguished only at marginal significance. { Data collected over a factor of $\sim4$ longer exposure time} are needed in order to measure $\sigma_s$ down to $\sim10\%$ accuracy and start conclusive probing of the $PDF$ shape. }

{Future generation of X-ray observatories will not only benefit from a factor of $\sim10$ higher sensitivity, but also from excellent angular and spectral resolution allowing minimisation of the systematic uncertainties due to contaminating signals. Even a modest exposure of $\sim 100$ ks with these instruments will be sufficient to probe the dynamic range of scales spanning $\sim1.5$ orders of magnitude, i.e. from few pc down to 0.1 pc. Further increase of this range will be challenging due to the effects of opacity and, possibly, finite duration of the flare (for which only an upper limit can currently be set).}

{Reconstruction of the $PDF$ shape across this range of scales in a uniform and unbiased manner will {shed light} on such key ingredients of the massive star formation paradigm as supersonic turbulence cascading and decay, transition to self-gravitating regions, seeding of star-forming cores and their feedback on the surrounding medium. In this regard, molecular clouds in the Central Molecular Zone are of particular importance given their extreme and highly dynamical environment as well as low star formation efficiency inferred for them from observations. } {Extension of the confidently reconstructed gas density $PDF$ to lower densities should also allow probing the connection of molecular clouds to the surrounding ISM, potentially including diagnostics of thermal stability and atomic-to-molecular transition of the dense gas in the CMZ environment (see Section \ref{s:global}).}

{A natural step forward will be spectral mapping of the reflected emission, in particular measurement of the fluorescent line centroid with very high (sub eV) accuracy. Although being very challenging, this might be feasible for the brightest clumps of the reflected emission, allowing direct measurement of the velocity field and its variation in the line-of-sight direction. X-ray data collected through sensitive and regular observations spread over $\sim10$ yrs should open a possibility for cross-correlation with velocity-resolved data on line emission from various molecular species and 3D velocity measurements provided by masers emission.}  

%
\section{Acknowledgements}
IK, EC and RS acknowledge partial support by the Russian Science Foundation grant~19-12-00369. {CF acknowledges funding provided by the Australian Research Council (Discovery Project DP170100603, and Future Fellowship FT180100495), and the Australia-Germany Joint Research Cooperation Scheme (UA-DAAD). {DS and SW acknowledge support by the German Science Foundation via CRC 956, sub-projects C5 and C6. SW further acknowledges support by the ERC Starting Grant RADFEEDBACK (grant no. 679852).} We further acknowledge high-performance computing resources provided by the Leibniz Rechenzentrum and the Gauss Centre for Supercomputing (grants~pr32lo {and pr94du}) and by the Australian National Computational Infrastructure (grant~ek9).} 







\newpage
\appendix

\section{Impact of contaminating sources of X-ray emission}
\label{s:appa}

{
Thanks to shortness of the illuminating flare and rich internal structure of the molecular gas, the observed reflected X-ray emission is characterised by prominent spatial and temporal variations \citep{2013A&A...558A..32C,2017MNRAS.465...45C}. In combination with its very distinct spectral shape, this allows one to confidently single out the reflected emission from various contaminating signals \citep[e.g.][]{2017MNRAS.468..165C}. 
} 

{
The contaminating signals can be broadly divided into two groups: diffuse (including apparently diffuse emission of unresolved faint point sources) and point-like, viz. resolved foreground and background (both Galactic and extra-galactic) point sources. While the former affects mostly the low surface brightness end of the distribution function, the contribution of the latter might be an issue for the statistics in the high surface brightness tail. Below we describe corresponding filtering techniques and the expected level of residual distortions inherent to data.
}

{There is also inevitable contamination of the primary reflection signal by the secondary emission arising due to scattering of the primary emission in the same cloud. We discuss impact of this contamination and possibilities of its filtering ({and even the usage} for more intricate density diagnostics). Finally, we comment on the implications of the multi-flare scenario, when emission from several illumination fronts might contribute to the reflection signal integrated over line-of-sight.}

\subsection{Diffuse emission}

\begin{figure}
\includegraphics[width=1.0\columnwidth,viewport=30 180 600 700]{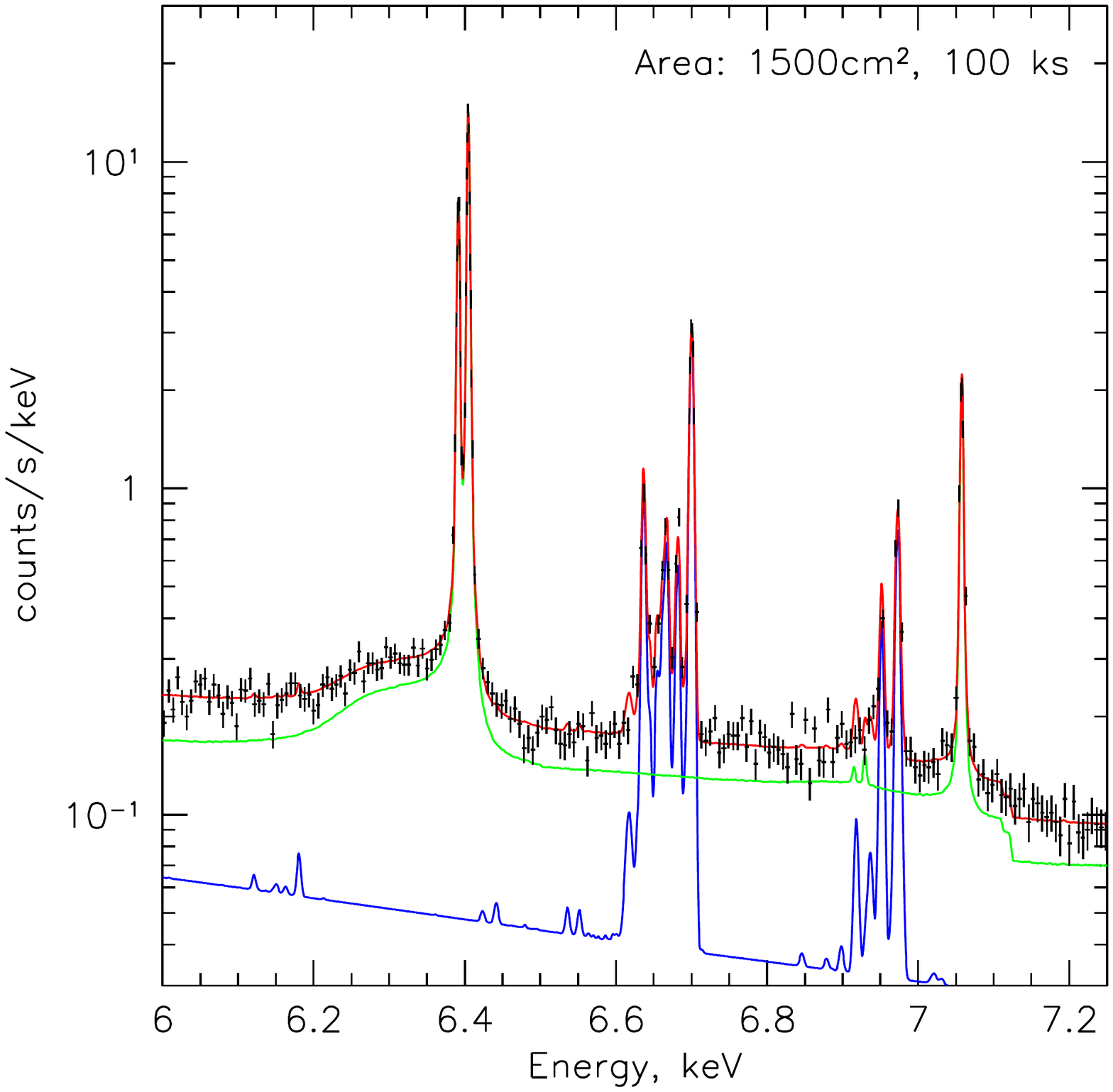}
\caption{
{Simulation of the high resolution 6-7.3 keV spectrum {(black crosses)} which will be obtained with future X-ray observatories having $\sim 1500\;{\rm cm^2}$ effective area and $\sim 3$~eV energy resolution after a 100-ks-long observation of a typical bright clump of reflected emission. {The green line shows the model of the reflected emission, the blue curve corresponds to the hot thermal plasma emission (schematically illustrating the contribution of unresolved compact sources), the red line shows their sum.
}
} 
\label{f:comp}
}
\end{figure}

\begin{figure}
\includegraphics[width=1.\columnwidth,viewport=50 210 580 690]{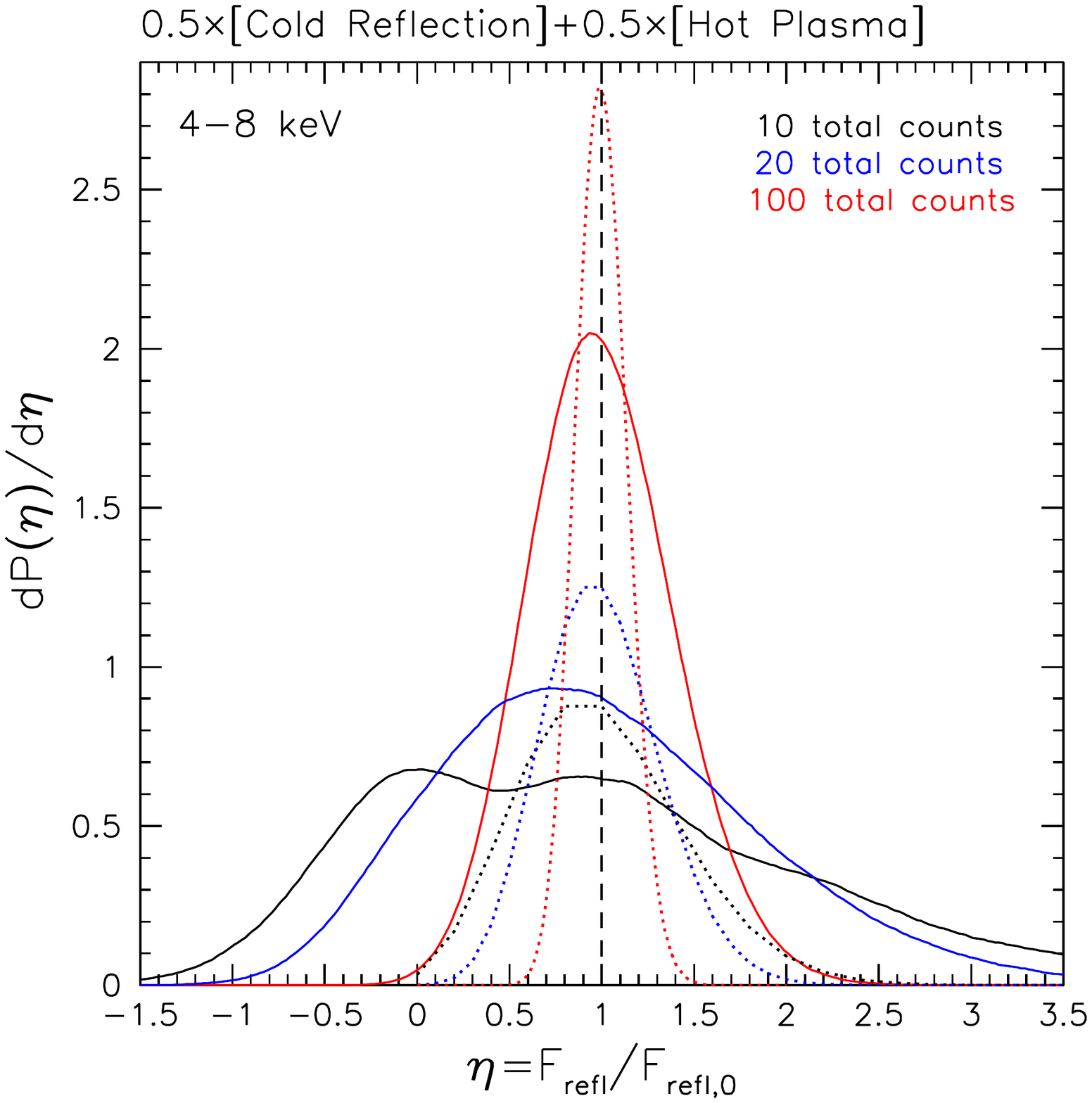}
\caption{
{
Simulated distribution of ratio of the recovered reflected component amplitude $F_{\rm refl}$ to the actual amplitude $F_{\rm refl,0}$ (solid curves) after spectral decomposition technique applied on an even mixture of the X-reflection and hot (T$\sim$6 keV) optically thin plasma emission in 4-8 keV energy band (given the spectral response function of \textit{Chandra} X-ray observatory). Black, blue and red curves correspond to 10, 20, and 100 total detected counts in this band, respectively. The dotted curves demonstrate distributions that would result from purely Poisson noise in the reflection component.}
}
\label{f:histcompsep}
\end{figure}

{
In addition to well-localised regions of thermal and non-thermal genuinely diffuse X-ray emission \citep[e.g. from shock-heated ISM due to young supernova remnants and, possibly, recent episodes of Sgr A* Myr-long activity,][]{2004ApJ...613..326M,2013MNRAS.434.1339H,2015MNRAS.453..172P,2016ApJ...825..132H,2018PASJ...70R...1K,2019Natur.567..347P}, bulk of apparently diffuse X-ray emission comes from the unresolved population of compact sources, mostly accreting white dwarfs \citep{2009Natur.458.1142R,2009ApJS..181..110M}. 
}

{
This emission is not only spatially smooth and non-variable, but it is also characterised by a distinct spectral shape, {4-8 keV part of} which is well approximated by emission model for hot optically thin plasma featuring prominent lines of highly ionised iron at 6.7 and 6.96 keV (see the blue line in Fig. \ref{f:comp}). On average, the surface brightness of this component is comparable to the mean surface brightness of the reflected emission, so its contribution has to be taken into account for the major part of the observed region (excluding the brightest knots of the reflected emission where it can be safely neglected).
}

{
In order to do that, we use the linear spectral decomposition method that relies {on evaluating the}  ``scalar product'' between the characteristic spectral templates of the emission components and {the spectrum} detected inside each individual pixel, {and then solving for best-fitting weights for each component.}
Since the spectral templates for X-ray reflection and hot plasma emission are not fully ``orthogonal'' (i.e. their ``scalar product'' is non-zero), {the resulting distribution is broader than that of the total flux combined, especially when the number of detected counts is small.} 
}{
Our tests of this procedure performed on data simulated with spectral response functions of \textit{Chandra} and \textit{XMM-Newton} X-ray observatories show that $\sim20$ counts detected in the 4-8 keV energy band are needed for unbiased estimation of the component amplitudes (as illustrated in Fig.~\ref{f:histcompsep} for the case of even components mixture). Given that equivalent width of the 6.4 keV line is $\sim1$ keV, this number simply guarantees that at least a few line counts get always detected.    
}{
In fact, the situation can be significantly improved by invoking additional information regarding expected amplitude for one of the components, e.g. based on its smooth (and even potentially predictable) spatial distribution. 
}

{
{Filtering of the diffuse contamination by the linear decomposition technique} leads a boost in statistical uncertainty of the flux estimation. Our tests show that this boost typically amounts to a factor of several compared to the pure Poisson uncertainty in one component (for an even component mixture and less than a few tens of total detected counts,{ as illustrated in Figure \ref{f:histcompsep})}. This should increase the impact of the Eddington bias on the $PDF$ reconstruction above the purely Poissonian expectation (see Appendix \ref{s:appb}). { The resulting distortion can be readily modelled given the actually measured amplitude of the contaminating signal in any particular region and kept under control.}
}

{
Of course, usage of two fixed spectral templates introduces some systematic uncertainty as well, which should be controlled by keeping the level of model complexity adequate to the available quality of the data (in order to avoid overfitting and resulting bias in the results and dramatic underestimation of the uncertainties). Future generations of X-ray observatories, featuring high sensitivity and high spectral resolution at energies around 6 keV, will allow us to minimise the impact of these systematic biases thanks to the possibility of using narrower spectral bands, e.g. centred on the iron fluorescence line at 6.4 keV (see Fig. \ref{f:comp} for the illustration), to improve quality of the component separation.
}

\subsection{Point sources}
{
Individual point sources that have spectral shape not too dissimilar from the spectral shape of X-ray reflection (like Galactic X-ray binaries or background active galactic nuclei) can potentially mimic the brightest knots of molecular gas. 
}{
Contamination by bright persistent compact sources can be suppressed by identifying persistent sources based on data of archival observations of the same region. Indeed, a compact gaseous clump (smaller than the available data resolution) can not be bright in the reflected emission longer than the duration of the flare, i.e. less than a few years (see a short discussion of the multi-flare scenario below).
}

{
For few transient sources that happened to be bright only in a particular observation, one can use a ``difference'' spectrum, by subtracting an archival spectrum from the same region. This approach removes all steady components and simplifies the classification of the variable emission based on the spectral shape. This approach has already been used to prove that the variable emission is indeed due to reflection \citep[see][]{2017MNRAS.468..165C}. 
}

{
Finally, the expected number of background sources can be estimated based on the well measured source count statistics in deep Galactic and extragalactic fields\citep[e.g.][]{2009Natur.458.1142R}. This should provide a statistical handle on possible number of unaccounted pointed sources and allow one to constrain their impact on the high-density tail of the reconstructed density $PDF$. 
}

\subsection{Second scatterings}
\label{ss:second}

\begin{figure}
\includegraphics[width=1.0\columnwidth,viewport=10 190 600 670]{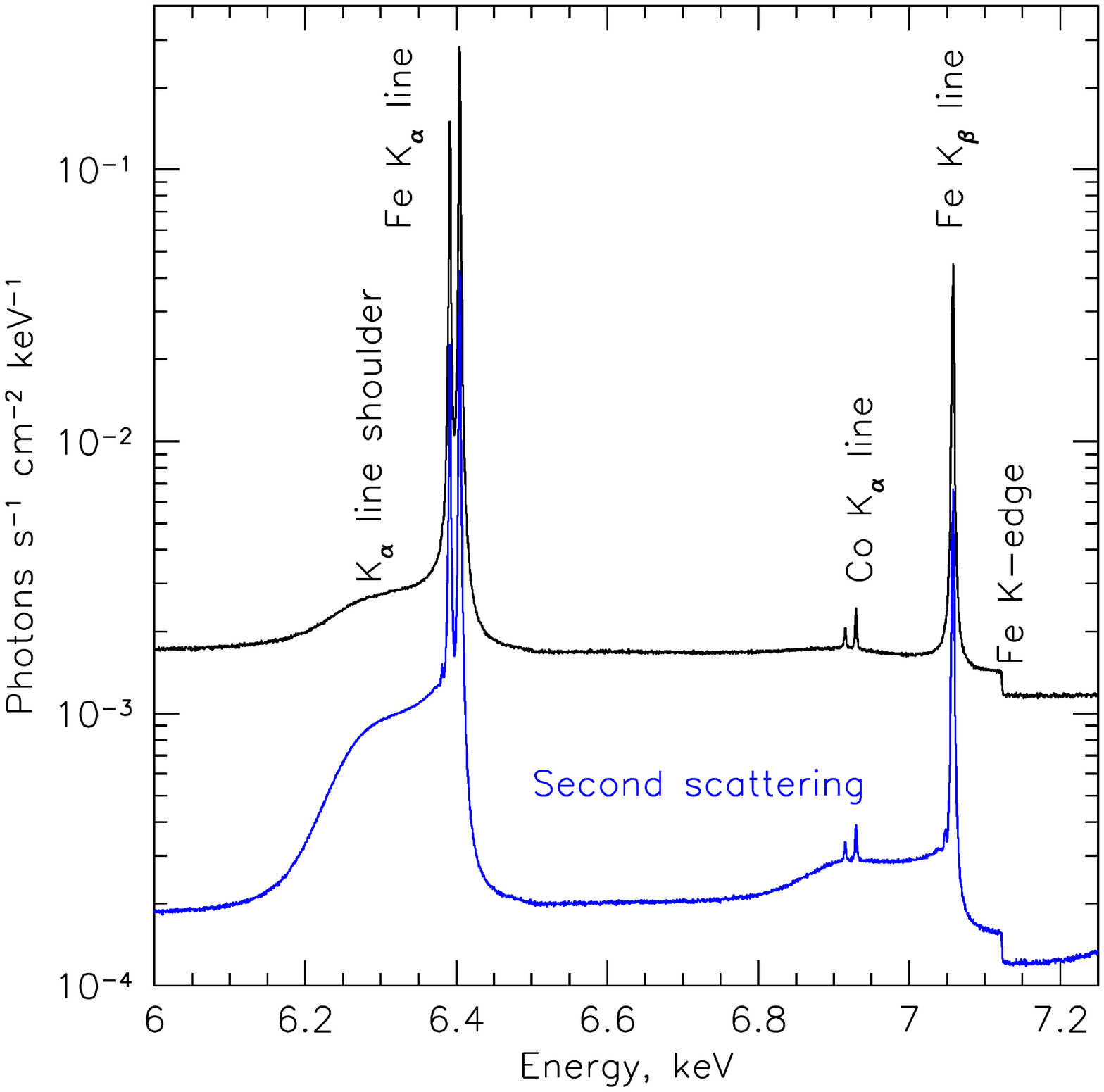}
\caption{
{Reflection spectrum from a slab of cold gas with the Thomson optical depth $\tau_T=0.2$ illuminated by a power-law X-ray continuum. The black line shows the total emission, the blue line - contribution of the doubly-scattered component. The main spectral features (i.e. fluorescent lines with corresponding Compton `shoulders' and absorption edges) are highlighted. The doubly-scattered emission can be observed even after the primary photons have already left the cloud.}
} 
\label{f:shoulder}
\end{figure}

{Scattering of the reflected emission in the same gas gives rise to the secondary emission, which is however typically much fainter due to small Thomson scattering depth of the cloud $\tau_{c}\ll1$. Indeed, for a compact gas clump entirely illuminated by the flare, the total reflected emission should be similar to the one shown in Fig.~\ref{f:shoulder} with the flux in the secondary component approximately {a factor $\tau_{c}$} smaller than the primary component. In particular, this is the case for the flux in the Compton `shoulder'  $F_{s}$ compared to the flux in its parent narrow fluorescent line $F_{l}$ (see Fig.~\ref{f:shoulder}).}

{For a larger cloud, whose light-crossing time is longer than the duration of the flare, the  strength of the secondary component should grow steadily with time as the flare front propagates across it. $F_s$ reaches the maximum value of $\tau_{c}\times F_{l}$ by the time when the front is about to leave the cloud. Once the front leaves the cloud, the ratio ${F_{s}}/{F_{l}}$  jumps to a level of order unity - an unambiguous spectral signature of the secondary emission \citep{1993ApJ...407..752C,1998MNRAS.297.1279S,2016A&A...589A..88M}. {Now, the iron 6.4~keV line complex is composed of the newly generated fluorescent line photons as well as the scattered line and line shoulder photons generated prior to the second scattering (see Fig.~\ref{f:shoulder}).} This process repeats with every next scattering and leads to an $N$-fold increase of the line complex equivalent width after $N$ scatterings. However, despite the increase of the equivalent width, the intensity of the multiply-scattered emission goes down, being proportional to $\tau_c^N$ for small $\tau_{c}$ and strongly diminished by photoelectric absorption during every scattering cycle for larger $\tau_c$.}

The life-time of the secondary component is set by {the longest of two time-scales - the light-crossing time of the cloud and the light-crossing time of the scattering environment on larger scales.  Therefore, the doubly-scattered (or multiply-scattered) emission probes the mean gas density of the entire region and, possibly, low-density extended envelopes of molecular clouds connecting them with the ambient ISM. In this case the flux of the doubly-scattered emission will be proportional to $\tau_c\tau_e$, where $\tau_e$ is the characteristic optical depth of the already illuminated extended region. Given the specific signatures of the second scattering, discussed above \citep[see also Fig.9  in ][]{1998MNRAS.297.1279S}, it should be possible to distinguish the doubly-scattered component from, e.g., the scattered emission from the gas having increased abundance of iron, especially with the anticipated high spectral resolution of future X-ray missions. Monitoring light curves of selected compact substructures and detecting variations of the 6.4 keV complex shape is therefore a very promising route towards characterisation of the gas distribution in a much larger volume.{
Observations of the long-living secondary emission from the clouds that are bright at the moment or were bright in the recent past (e.g. the giant molecular complex Sgr B2) with future X-ray missions will be particularly valuable in this regard.
}
}

\subsection{Impact of multiple illuminating flares}
{
Finally, we comment on implications of a possible scenario in which echoes of multiple flaring events are currently observed in X-ray reflection on the clouds in the CMZ \citep{2013A&A...558A..32C,2018A&A...610A..34C,2018A&A...612A.102T}. In this scenario, the reflected X-ray emission from a thin slice of illuminated molecular gas might be contaminated by reflected emission of a different slice illuminated by another flare.}

{
The observed strong variability of the brightest knots of X-ray emission and consistency of its statistical properties with statistical properties of the spatial variations \citep{2017MNRAS.465...45C} imply that there is likely no substantial projection overlap of multiple illuminated slices, at least for some of the illuminated clouds. If the illuminated slices actually belong to the same cloud, then this scenario is simply similar to a scenario with a single flare of longer duration, at least for the scales probing of which is not strongly affected by de-correlation due to spatial separation of the slices (very roughly all scales below the slices separation length).  
}

\section{Eddington bias for Poisson log-normal distribution}
\label{s:appb}

{
The situation of log-normal parent distribution measured with superimposed Poisson noise takes place on many occasions in bio-statistical studies, so it has been well explored \citep[e.g.][]{Bulmer1974}. The possibility of using the saddle point approximation for evaluating the convolution integral has been proposed in that context by \cite{Izsak2008}. We outline the method and present the resulting concise analytic expression below. 
}

{
Probability of detecting $C$ counts in a pixel of X-ray map with log-normal underlying distribution of the surface brightness is written as
\begin{equation}
P(C)=\int_\lambda\frac{e^{-\lambda}}{C!}\lambda^{C} p(\lambda)d\lambda.
\end{equation}
where $\lambda$ is the expected number of counts for a given surface brightness value $I_X$, and $p(\lambda)$ is its original distribution function. The former can be expressed through the expected number of counts for the mean surface brightness $I_0$ as $\lambda=C_0 I_X/I_{0}=C_0 e^{s}$. 
}

{
Having defined $\xi=\ln\lambda$ and $\xi_0=\ln C_0$, one has $s=\xi-\xi_0$, so that the log-normal distribution gets form of  
\begin{equation}
p(\lambda)d\lambda=p_\xi d\xi=\frac{1}{\sqrt{2\pi\sigma_s^2}}\exp \left[-\frac{(\xi-\xi_0+\frac{\sigma_s^2}{2})^2}{2\sigma_s^2}\right]d\xi.
\end{equation}
Hence
\begin{equation}
P(C)=\frac{1}{C!\sqrt{2\pi\sigma_s^2}}\int_{-\infty}^{\infty} \exp\left[\xi C-e^{\xi}-\frac{(\xi-\xi_0+\frac{\sigma_s^2}{2})^2}{2\sigma_s^2}\right]d\xi.
\label{eq:pc}
\end{equation}
}

{
Let us define 
\begin{equation}
\varphi(\xi)=\xi C-e^{\xi}-\frac{(\xi-\xi_0+\frac{\sigma_s^2}{2})^2}{2\sigma_s^2},
\end{equation}
so that equation \ref{eq:pc} takes form $P(C)=A\int e^{\varphi(\xi)}d\xi$.
}

{
Given that $\varphi(\xi)\rightarrow -\infty$ for $\xi \rightarrow \pm\infty$, the saddle point method can be used to approximately evaluate this integral by taking into account only the region close to the global maximum of  $\varphi(\xi)$. To localise it, we solve the equation
\begin{equation}
\varphi\prime(\xi)
=C+\frac{\xi_0-\frac{\sigma_s^2}{2}}{\sigma_s^2}-e^{\xi}-\frac{\xi}{\sigma_s^2}=0.
\end{equation}
}

{
Due to the presence of the exponent, the solution can be found via iterative Taylor expansions starting from 
\begin{equation}
\xi_1=\ln \left[C+\frac{\xi_0-\frac{\sigma_s^2}{2}}{\sigma_s^2}\right]  
\label{eq:xi1}
\end{equation}
and proceeding in leftward direction:
\begin{equation}
\xi_{2}=\xi_1-\Delta_{\xi,1}, ~~\mathrm{with}~ \Delta_{\xi,1}=\frac{\xi_1}{C\sigma_s^2+\xi_0+1}
\label{eq:xi2}
\end{equation}
and
\begin{equation}
\xi_{m}=\xi_2-\Delta_{\xi,2}, ~~\mathrm{with}~ \Delta_{\xi,2}=\frac{\xi_2-\xi_0-C\sigma_s^2+\sigma_s^2e^{\xi_2}}{\sigma_s^2e^{\xi_2}+1}.
\label{eq:xim}
\end{equation}
According to our tests, these two iterations provide sufficient accuracy in all practically relevant cases.
}

{
Since $\varphi\prime(\xi)=0$ for $\xi=\xi_{m}$, in its vicinity one has 
\begin{equation}
e^{\varphi(\xi)}\approx e^{\varphi(\xi_{m})}\times  \exp{\left[{-\frac{(\xi-\xi_{m})^2}{2\sigma_{\xi}^2}}\right]}, 
\end{equation}
where $\sigma_{\xi}=1/\sqrt{-\varphi\prime\prime(\xi_m)}=1/\sqrt{e^{\xi_m}+{1}/{\sigma_s^2}}$.
}

{
This approximation allows the integral $\int e^{\varphi(\xi)}d\xi$ to be taken analytically, resulting in very compact expression:
\begin{equation}
P(C)=\frac{e^{\xi_m}}{C!\,\sqrt{e^{\xi_{m}}\sigma_s^2}+1}.
\label{eq:pcf}
\end{equation}
The set of equations \ref{eq:xi1}-\ref{eq:xim} and \ref{eq:pcf} presents the fully-analytic approximate solution of the problem. { This solution might be used as a starting point for further modelling taking into account presence of (uncorrelated) background noise and effects of spectral component separation technique (see Appendix \ref{s:appa}), or more complicated shape of the underlying distribution function (as for instance bi-modal log-normal distributions presented in Section \ref{s:global}).} 
}

\bsp	
\label{lastpage}
\end{document}